\documentclass[twocolumn]{jpsj2} 
\title{Dynamical CPA and Tight-Binding LMTO Approach to Correlated
Electron System}

\author{Yoshiro \textsc{Kakehashi}\thanks{E-mail address:
yok@sci.u-ryukyu.ac.jp, to be published in Phys. Soc. Jpn. {\bf 77} No.9 
(2008)},
Takafumi \textsc{Shimabukuro}, Toshihito \textsc{Tamashiro}, \\ 
and Tetsuro \textsc{Nakamura} 
}

\inst{
Department of Physics and Earth Sciences,
Faculty of Science, University of Ryukyus, \\
1 Senbaru, Nishihara, Okinawa, 903-0213, Japan
}

\abst{
Dynamical Coherent-Potential Approximation (CPA) to correlated electrons
has been extended to a system with realistic Hamiltonian which consists 
of the first-principles tight-binding Linear Muffintin Orbital (LMTO) 
bands and intraatomic Coulomb interactions.
Thermodynamic potential and self-consistent equations for Green
function are obtained on the basis of the
functional integral method and the harmonic approximation which neglects
the mode-mode couplings between the dynamical potentials with different
frequency.  Numerical calculations have been performed for Fe and Ni
within the 2nd-order dynamical corrections to the static approximation.
The band narrowing of the quasiparticle states and the 
6 eV satellite are obtained for Ni at finite temperatures.
The theory leads to the Curie-Weiss law for both Fe and Ni.
Calculated effective Bohr magneton numbers are 3.0 $\mu_{\rm B}$ for Fe 
and 1.2 $\mu_{\rm B}$ for Ni, explaining the experimental data. 
But calculated Curie temperatures are
2020 K for Fe and 1260 K for Ni, being still overestimated
by a factor of two as compared with the experimental ones. 
Dynamical effects on electronic and magnetic properties are discussed by
comparing with those in the static approximations. 
}

\kword{dynamical CPA, correlated electrons, tight-binding Linear
Muffintin Orbital method, iron, nickel, Curie temperature, effective
Bohr magneton number, quasiparticle states}

\begin{document}
\maketitle

\section{Introduction} 
Understanding of electronic and magnetic properties of the system with
intermediate strength of Coulomb interactions has been a challenging
problem over half a century in condensed matter physics, because simple
theoretical approaches are not applicable to the system in spite of 
the fact that many
intriguing phenomena are found there~\cite{fulde95,imada98,kake04,fulde06}.  
Iron and nickel are considered to be an example of such systems.  
These metals in fact show the properties of both the weakly-
and the strongly- correlated electrons.
Photoemission data, for example, show 
the existence of metallic $d$ bands~\cite{eastman80,schafer05} 
and the Sommerfeld coefficients in the $T$-linear specific heats show
rather large values (5-7 mJ/K${}^{2}$mol) as compared with those of the
noble metal systems~\cite{bozorth68}.  
The quasiparticle band widths however are found to be narrower
than the results of usual band calculations and a satellite peak
is observed at 6 eV below the Fermi level in Ni~\cite{himpsel79}, 
which are not able to be explained 
by a simple band theory~\cite{moruzzi95}, 
suggesting rather strong electron correlations in these systems.
The same features are found in the magnetic properties.  Noninteger
values of the ground-state magnetization in Fe and Ni are well explained
by the band theory~\cite{moruzzi78}, 
while the paramagnetic susceptibilities of these
systems follow the Curie-Weiss law and their effective Bohr magneton
numbers are close to those expected from the local-moment 
model~\cite{bozorth68}.  
Large specific heats near the Curie temperature $T_{\rm C}$ 
are also well explained by the same model.  

The magnetic and electronic properties in the intermediate regime of
Coulomb interactions have been traditionally explained by interpolation
theories between the weak and strong Coulomb interaction limits.  
Cyrot~\cite{cyrot72}
proposed an interpolation theory on the basis of the functional integral
method which transforms an interacting electron system into an
independent electron system with time-dependent random charge and
exchange fields.  He showed that the static and saddle-point
approximations to the functional integral scheme can explain the 
local-moment vs. itinerant behavior in magnetism of transition metals, 
as well as the metal-insulator transition.

Hubbard~\cite{hub79} and Hasegawa~\cite{hase79} independently 
developed a single-site spin fluctuation theory.  
They adopted a high temperature approximation ({\it i.e.}, the static 
approximation), and treated the random charge
and exchange potentials by making use of the coherent potential
approximation (CPA).  The theory qualitatively described the
magnetization vs. temperature curves, the Curie temperature, as well as
the Curie-Weiss susceptibility in Fe and Ni.  
The theory, however, reduces to the
Hartree-Fock approximation at zero temperature because it relies on 
the static approximation.  This means that the theory does not take 
into account the ground-state electron correlations as
discussed by Gutzwiller~\cite{gutz63}, Hubbard~\cite{hub63}, 
and Kanamori~\cite{kana63}.  Furthermore, the
quasiparticle bands and the satellite peak do not appear in the theory
using the static approximation.

We proposed the dynamical CPA which fully takes into account
the electron correlations within the single-site approximation, and
clarified the qualitative features of dynamical effects using a
Monte-Carlo sampling method~\cite{kake92}.  
More recently, we developed analytic
method to the dynamical CPA~\cite{kake02}, 
adopting the harmonic approximation
(HA)~\cite{amit71}.  
The latter is based on the neglect of the mode-mode couplings
between dynamical potentials in solving an impurity problem in an
effective medium.  
The HA interpolates between the weak Coulomb interaction
limit and the atomic limit.  Especially it describes the Kondo behavior
quantitatively in the strong correlation limit~\cite{dai91}.  
We showed within the single band model that the
dynamical CPA+HA yields the band narrowing of quasiparticle states and
the satellite peak in Fe and Ni, which were not explained by the early
theories with use of the static approximation.  The theory was however 
based on the single-band Hubbard model.  Quantitative calculations
of transition metals and alloys with use of the realistic Hamiltonian 
has not yet been made even within the single-site approximation.

In the present paper, we extend the dynamical CPA to the multi-band case
adopting the first-principles tight-binding (TB) linear muffintin 
orbitals (LMTO) method~\cite{ander75,ander94}.  
The modern band theory is based on the density functional theory (DFT) 
which allows us to express the ground-state energy as a functional of 
the spin and charge densities of the system~\cite{par89}.
In the local density approximation (LDA)~\cite{barth72}, 
one approximates the energy
functional with the energy function of the density.  The LDA 
exchange-correlation 
potentials obtained from the electron gas system have much
simplified the electronic band-structure calculations in solids.  The
TB-LMTO method allows us to construct the first-principles
tight-binding one electron Hamiltonian, and to calculate the LDA band
structure.   We adopt the
TB-LMTO Hamiltonian to describe the noninteracting part of the
Hamiltonian, and take into account the intraatomic Coulomb and exchange
interactions between $d$ electrons which are dominant among
electron-electron interactions.

Similar theoretical approach has been developed in the problem of the
metal-insulator transition in infinite dimensions~\cite{georges96}.
The approach called the dynamical mean field theory (DMFT) 
is equivalent to the dynamical CPA, as we have shown 
recently~\cite{kake02-2}.
The present theory therefore should be equivalent in principle 
to the DMFT combined with the LDA+U scheme in the band 
theory~\cite{anisimov97}. 
The merits of the present approach may be summarized as follows. 
(1) The dynamical CPA can treat the transverse spin
fluctuations for arbitrary $d$ electron number at finite temperatures ,
while the standard DMFT combined with the quantum Monte-Carlo method 
(QMC) cannot treat them because 
it is based on the Ising-type Hubbard-Stratonovich 
transformation~\cite{hirsch89}. 
Because of the reason, the DMFT calculations for Fe 
and Ni have been performed so far without taking into account the 
transverse spin fluctuations at finite temperatures~\cite{lich01}. 
(2) The HA which we adopted to solve the impurity problem is an analytic
approach from the high-temperature limit.  
The approach is suitable for understanding the finite-temperature
magnetism because the zero-th approximation to the HA describes
the magnetic properties much better than the Hartree-Fock one.  
There is no corresponding
approach in the DMFT. (3) Because of the analytic theory,
we can calculate the excitation spectra up to the
temperatures much lower than those calculated by the QMC, using 
the Pad\'e numerical analytic continuation method~\cite{vidberg77}.

In the following section, we introduce a TB-LMTO Hamiltonian with
intraatomic Coulomb interactions.  In \S 3, we
formulate the dynamical CPA to the realistic Hamiltonian on the basis of
the functional integral technique~\cite{morandi74}.  Applying a
generalized Hubbard-Stratonovich transformation~\cite{hub59} 
to the free energy, we transform the
interacting electrons into an independent electron system with time
dependent random fields.  Introducing an effective medium into the time
dependent Hamiltonian, we will
make a single-site approximation.  We determine the medium solving a
self-consistent equation, called the CPA equation~\cite{ehren76}.
In \S 4, we adopt the HA to calculate the dynamical part of the free
energy, and derive the analytic expressions of the free energy, 
the dynamical CPA equation, and other thermodynamic quantities.
In \S 5, we present the numerical results of calculation for Fe and Ni.
The calculations have been performed by using the second-order 
dynamical CPA ({\it i.e.}, the dynamical CPA+HA within the second-order
dynamical corrections).
We explain the band narrowing of the quasiparticle states, the
incoherent satellite peak at 6 eV below the Fermi level in Ni.
We also present the results of calculations for the magnetization
vs. temperature curve, the paramagnetic susceptibility following the
Curie-Weiss law, and the amplitude of local moments.
We clarify the quantitative aspects of the theory comparing with the
experimental data, and examine 
the dynamical effects on various quantities comparing the dynamical
results with those in the static approximation.
The last section 6 is devoted to summarize the dynamical CPA and
TB-LMTO Hamiltonian approach, as well as the dynamical effects in Fe 
and Ni.
\vspace{10mm}

\section{TB-LMTO Hamiltonian}
We adopt in the present paper the first-principles TB-LMTO 
method~\cite{ander94} 
to construct a realistic many-body Hamiltonian.
In this case, atomic basis function with orbital $L$ on site $i$, 
$\chi_{iL}(\mbox{\boldmath $r$}-\mbox{\boldmath $R$}_{i})$,
are constructed from a muffintin atomic orbital 
$\varphi_{iL}(\mbox{\boldmath $r$}-\mbox{\boldmath $R$}_{i})$
on site $i$ with an atomic level $E_{\nu i L}$, and a tail function 
outside the muffintin potential $h_{jL^{\prime}iL}^{\alpha}$ as
\begin{eqnarray}
\chi_{iL}(\mbox{\boldmath $r$}\!-\!\mbox{\boldmath $R$}_{i})  = 
\varphi_{iL}(\mbox{\boldmath $r$}\!-\!\mbox{\boldmath $R$}_{i}) \! + \!
\sum_{jL^{\prime}} \dot{\varphi}_{jL^{\prime}}^{\alpha}
(\mbox{\boldmath $r$}\!-\!\mbox{\boldmath $R$}_{j})
h_{jL^{\prime}iL}^{\alpha} ,
\label{chi}
\end{eqnarray}
\begin{eqnarray}
\dot{\varphi}_{iL}^{\alpha}
(\mbox{\boldmath $r$}-\mbox{\boldmath $R$}_{i}) = 
\dot{\varphi}_{iL}(\mbox{\boldmath $r$}-\mbox{\boldmath $R$}_{i}) +
\varphi_{iL}(\mbox{\boldmath $r$}-\mbox{\boldmath $R$}_{i})
o_{iL}^{\alpha} \ .
\label{phidot}
\end{eqnarray}
Here the wave function in the interstitial region has been neglected
because of the atomic sphere approximation.  The atomic wave function 
$\varphi_{iL}(\mbox{\boldmath $r$}-\mbox{\boldmath $R$}_{i})$ and
its energy derivative $\dot{\varphi}_{iL}(\mbox{\boldmath $r$})$ 
are defined by 
$\varphi_{iL}(\mbox{\boldmath $r$})=
\phi_{iL}(E_{\nu iL},r)Y_{L}(\hat{\mbox{\boldmath $r$}})$ and 
$\dot{\varphi}_{iL}(\mbox{\boldmath $r$})=
\dot{\phi}_{iL}(E_{\nu iL},r)Y_{L}(\hat{\mbox{\boldmath $r$}})$,
where $Y_{L}(\hat{\mbox{\boldmath $r$}})$ is the cubic harmonics with 
$L=(l,m)$, $l$ being the azimuthal quantum number and $m$ being an
orbital index for $l$.  $\phi_{iL}(E,r)$ is obtained by solving 
the radial Schr\"odinger equation with energy $E$.  
The energy $E_{\nu iL}$ is chosen to be the center of gravity below the
Fermi level for each orbital.  The atomic orbitals $\{ \varphi_{iL} \}$ 
are normalized in the atomic sphere as 
$\langle \varphi_{iL} | \varphi_{iL^{\prime}} \rangle = 
\delta_{LL^{\prime}}$.
The tail coefficients $h^{\alpha}_{jL^{\prime}iL}$ in eq. (\ref{chi}) are
determined in such a way that the orbital $\chi_{iL}$ is continuous and
differentiable on the sphere boundary at each sphere.   
The coefficient $o^{\alpha}_{iL}$ in eq. (\ref{phidot}) is determined 
so that the orbital $\chi_{iL}$ is well localized.
We adopt here the nearly orthogonal representation 
({\it i.e.} $o^{\alpha}_{iL}=0$), in
which the orbitals $\chi_{iL}$'s become orthogonal up to second order in 
$h_{iLjL^{\prime}}$.  The TB-LMTO Hamiltonian matrix is then written as
\begin{eqnarray}
H_{iLjL^{\prime}} & = & \langle \chi_{iL} | 
\left( -\nabla^{2} + v(\mbox{\boldmath $r$}) \right) | \chi_{jL} \rangle
\nonumber \\ 
& = & \epsilon_{iL}\delta_{ij}\delta_{LL^{\prime}} + t_{iLjL^{\prime}} .
\label{hij}
\end{eqnarray}
Here $v(\mbox{\boldmath $r$})$ is a LDA potential, $\epsilon_{iL}$ is an
atomic level, and $t_{iLjL^{\prime}}$ is a transfer integral between
orbitals $\chi_{iL}$ and $\chi_{jL^{\prime}}$.

When we adopt the density functional theory, the one-electron
Hamiltonian (\ref{hij}), especially the atomic level $\epsilon_{iL}$, 
contains the effects of strong intratomic
Coulomb interactions in general.  According to the LDA+U interpretation 
by Anisimov {\it et. al.}~\cite{anisimov97-2},
the atomic level $\epsilon^{0}_{iL}$ for noninteracting system is
obtained from the relation,
\begin{eqnarray}
\epsilon^{0}_{iL} = \dfrac{\partial E_{\rm LDA}}{\partial n_{iL\sigma}} 
- \dfrac{\partial E^{\rm U}_{\rm LDA}}{\partial n_{iL\sigma}} \ .
\label{e0}
\end{eqnarray}
Here $n_{iL\sigma}$ is the charge density at the ground state, 
$E_{\rm LDA}$ is the ground-state energy in the LDA, and 
$E^{\rm U}_{\rm LDA}$ is a LDA functional to the intraatomic Coulomb 
interactions.
Among various forms of $E^{\rm U}_{\rm LDA}$, we adopt the Hartree-Fock
type form~\cite{anisimov93} 
since we consider here an itinerant electron system 
where the ratio of the Coulomb interaction to the $d$ band width 
is not larger than one.
\begin{eqnarray}
E^{\rm U}_{\rm LDA} & = & \dfrac{1}{2} \sum_{j} \sum_{mm^{\prime}\sigma} 
\overline{U} n_{jd} n_{jd}   \nonumber \\
& &  \hspace{5mm}
+ \dfrac{1}{2} \sum_{j} {\sum_{mm^{\prime}}}^{\prime} \sum_{\sigma} 
(\overline{U} - \overline{J}) n_{jd}n_{jd} \ .
\label{eulda}
\end{eqnarray}
Here $n_{jd}=\sum_{m\sigma} n_{jlm\sigma}/2(2l+1)$ with $l=2$.
$\overline{U}$ and $\overline{J}$ are the orbital-averaged Coulomb and
exchange interactions defined by
\begin{eqnarray}
\overline{U} = \dfrac{1}{(2l+1)^{2}} \sum_{mm^{\prime}} U_{mm^{\prime}} \ ,
\label{ubar}
\end{eqnarray}
\begin{eqnarray}
(\overline{U} - \overline{J}) = 
\dfrac{1}{2l(2l+1)} {\sum_{mm^{\prime}}}^{\prime} 
( U_{mm^{\prime}} -  J_{mm^{\prime}} ) \ ,
\label{u-j}
\end{eqnarray}
where $U_{mm^{\prime}}$ and $J_{mm^{\prime}}$ are orbital dependent
intraatomic Coulomb and exchange integrals for $d$ electrons.
From eqs. (\ref{e0}) and (\ref{eulda}), we obtain the atomic level 
$\epsilon^{0}_{iL}$ for noninteracting system as
\begin{eqnarray}
\epsilon^{0}_{iL} & = & \epsilon_{iL} 
- \bigg[ \left( 1 - \dfrac{1}{2(2l+1)} \right) \overline{U} 
\nonumber \\
& &   \hspace{13mm}
- \dfrac{1}{2} \left( 1 - \dfrac{1}{(2l+1)} 
\right) \overline{J} \bigg] n_{d} \delta_{l2} \ .  \hspace{5mm}
\label{e02}
\end{eqnarray}
Note that $n_{d}$ denotes the total $d$ electron number per atom.

The Hamiltonian which we consider here can be written as
\begin{eqnarray}
\hat{H} = H_{0} + H_{1} .
\label{hhat}
\end{eqnarray}
The tight-binding Hamiltonian for noninteracting system $H_{0}$ 
is given by 
\begin{eqnarray}
H_{0} = \sum_{iL\sigma} (\epsilon^{0}_{iL} - \mu) \, 
\hat{n}_{iL \sigma} + \sum_{iL jL^{\prime} \sigma} t_{iL jL^{\prime}} \, 
a_{iL \sigma}^{\dagger} a_{jL^{\prime} \sigma} \ .
\label{h0}
\end{eqnarray}
Here we have introduced the chemical potential $\mu$ for the calculation
of the free energy.  $a_{iL \sigma}^{\dagger}$ 
($a_{iL \sigma}$) is the creation (annihilation) operator for an
electron with orbital $L$ and spin $\sigma$ on site $i$, and 
$\hat{n}_{iL\sigma}=a_{iL \sigma}^{\dagger}a_{iL \sigma}$ is a charge
density operator for electrons with orbital $L$ and spin $\sigma$ on
site $i$.  We have neglected the change of the transfer integrals due to
electron-electron interactions.

The interacting part $H_{1}$ in eq. (\ref{hhat}) 
consists of the intraatomic Coulomb interactions between $d$ electrons.
\begin{eqnarray}
H_{1} & = & \sum_{i} 
\Big[ \sum_{m} U_{0} \, \hat{n}_{ilm \uparrow} \hat{n}_{ilm \downarrow} 
\nonumber \\
& &   \hspace{-12mm}
+ {\sum_{m > m^{\prime}}} 
(U_{1}-\frac{1}{2}J) \hat{n}_{ilm} \hat{n}_{ilm^{\prime}} -
{\sum_{m > m^{\prime}}} J   
\hat{\boldsymbol{s}}_{ilm} \cdot \hat{\boldsymbol{s}}_{ilm^{\prime}} 
\Big] . \hspace{7mm}
\label{h1}
\end{eqnarray}
Here $U_{0}$ ($U_{1}$) and $J$ are the intra-orbital (inter-orbital)
Coulomb interaction and the exchange interaction, respectively.  
$\hat{n}_{ilm}$ ($\hat{\boldsymbol{s}}_{ilm}$) with $l=2$ is 
the charge (spin)
density operator for $d$ electrons on site $i$ and orbital $m$,  
which is defined by $\hat{n}_{ilm}=\sum_{\sigma} \hat{n}_{ilm\sigma}$
($\hat{\boldsymbol{s}}_{ilm} = \sum_{\alpha\gamma} 
a_{iL \alpha}^{\dagger} (\boldsymbol{\sigma}/2)_{\alpha\gamma} 
a_{iL \gamma} $), $\boldsymbol{\sigma}$ being the Pauli spin matrices.
\vspace{10mm}

\section{Functional Integral Approach and Dynamical CPA}
Thermodynamic properties of the system are calculated from the partition
function, which is given by 
\begin{eqnarray}
Z = {\rm Tr} \left[ {\cal T} \exp \left( - \int^{\beta}_{0} (H_{0}(\tau) 
+ H_{1}(\tau)) \right) \right] .
\label{pf}
\end{eqnarray}
Here $\beta$ is the inverse temperature, ${\cal T}$ denotes the
time-ordered product ($T$-product) for operators.  $H_{0}(\tau)$ 
($H_{1}(\tau)$) is the interaction representation of Hamiltonian 
$H_{0}$ ($H_{1}$).

The functional integral method is based on a Gaussian formula for the
Bose-type operators $\{ a_{\mu} \}$.
\begin{eqnarray}
{\rm e}^{\displaystyle \ \sum_{mm^{\prime}} 
a_{m} A_{mm^{\prime}} a_{m^{\prime}}} \hspace{-7mm} & = &  \hspace{-3mm}
\ \sqrt{\dfrac{{\rm det} A}{\pi^{M}}} \int [ \prod_{m} dx_{m} ]
\nonumber \\
& & \hspace{-15mm}
\ {\rm e}^{\displaystyle - \sum_{mm^{\prime}} (x_{m} A_{mm^{\prime}} 
x_{m^{\prime}} - 2a_{m} A_{mm^{\prime}}x_{m^{\prime}})} .  \hspace{5mm}
\label{gauss}
\end{eqnarray}
Here $A_{mm^{\prime}}$ is a $M \times M$ matrix, and 
$\{ x_{m} \}$ are auxiliary field variables. 
Discretizing the integral with respect to time in eq. (\ref{pf}), and 
applying the formula (\ref{gauss}) to the bose-type operators at each
time under the $T$-product, 
we obtain a functional integral form of the free energy
${\cal F}$ as
\begin{eqnarray}
e^{-\beta {\cal F}} & = & 
\int \left[ \prod_{i=1}^{N} \prod_{m=1}^{2l+1} 
\delta \boldsymbol{\xi}_{im}(\tau) \delta \zeta_{im}(\tau) \right]
Z^{0}(\boldsymbol{\xi}(\tau), \zeta(\tau))  \nonumber  \\ 
&  &  \hspace{-15mm} \times 
\exp \bigg[ - \frac{1}{4} \sum_{i} {\sum_{mm^{\prime}}}^{\prime} 
\!\!
\int^{\beta}_{0} d\tau \Big( \zeta_{im}(\tau) A_{imm^{\prime}} 
\zeta_{m^{\prime}}(\tau)   \nonumber \\
& &    \hspace{10mm}
+ \sum_{\alpha}^{xyz} \xi_{im\alpha}(\tau) 
B^{\alpha}_{imm^{\prime}} \xi_{im^{\prime}\alpha}(\tau) \Big) \bigg] ,
\label{free0}
\end{eqnarray}
\begin{eqnarray}
Z^{0}(\boldsymbol{\xi}(\tau), \zeta(\tau))
= {\rm Tr} \left( \!\! {\cal T}  \!
\exp \left[ - \!\! \int^{\beta}_{0} \!\!\! H(\tau, \boldsymbol{\xi}, -i\zeta) 
d\tau \right] \right) ,
\label{z0}
\end{eqnarray}
\begin{eqnarray}
H(\tau, \boldsymbol{\xi}, \!\! -i\zeta) \hspace{-4mm} & = & \hspace{-5mm}
\sum_{iL} \! \Big[ \Big( \! \epsilon^{0}_{iL} \!\! - \!\! \mu \!! - \!\! 
\frac{1}{2} \!\! \sum_{m^{\prime}} i A_{imm^{\prime}} \zeta_{im^{\prime}}(\tau)
\delta_{l2} \! \Big) \hat{n}_{iL}(\tau)  \nonumber \\
& & \!\!\!\!\!\!\!\!\!\!\!\!\!\!\!\!\!\!\!\!\!\!
- \sum_{\alpha} \left( \frac{1}{2} \sum_{m^{\prime}} 
B^{\alpha}_{imm^{\prime}} \xi_{im^{\prime}\alpha}(\tau) + 
h^{\alpha}_{im} \right)\delta_{l2} \hat{m}^{\alpha}_{iL}(\tau) \Big]
\nonumber \\
& &  \hspace{-10mm}
+ \sum_{iLjL^{\prime}\sigma} t_{iLjL^{\prime}} a^{\dagger}_{iL\sigma}(\tau)
a_{jL^{\prime}\sigma}(\tau) \ .
\label{htau}
\end{eqnarray}
Here $N$ is the number of sites, 
$\hat{\boldsymbol{m}}_{iL} = 2\hat{\boldsymbol{s}}_{iL}$, and 
$\zeta_{im}(\tau)$ ($\boldsymbol{\xi}_{im}(\tau)$) is an auxiliary
field being conjugate with $i\hat{n}_{iL}(\tau)$
($\hat{\boldsymbol{m}}_{iL}(\tau)$) for $l=2$. 
The functional integrals in eq. (\ref{free0}) are, for example, 
defined by
\begin{eqnarray} 
\int \! \left[ \! \prod_{m=1}^{2l+1} \!\! \delta \zeta_{im}(\tau) \! 
\right] \!\!
= \!\! \int  \!\! \left[ \! \prod_{n=1}^{N^{\prime}} \!\!
\sqrt{\! \dfrac{\beta^{2l+1} \! {\rm det} A_{i}}{(4\pi)^{2l+1}}} \!\!
\prod_{m=1}^{2l+1} \!\!
\dfrac{d\zeta_{im}(\tau_{n})}{\sqrt{N^{\prime}}} \right]  , \!
\label{func}
\end{eqnarray}
where $2l+1$ in the square roots denotes the number of $d$ orbitals 
({\it i.e,} $2l+1=5$).
${\rm det} A_{i}$ is the determinant of 
the $(2l+1) \times (2l+1)$ matrix $A_{imm^{\prime}}$. 
$\tau_{n}$ denotes the $n$-th time when 
the time interval $[0,\beta]$ is divided into $N^{\prime}$ segments.
The matrices $A_{imm^{\prime}}$ and 
$B^{\alpha}_{imm^{\prime}}$ ($\alpha=x,y,z$) are defined as
\begin{eqnarray} 
A_{imm^{\prime}} = U_{0}\delta_{mm^{\prime}} 
+ (2U_{1} - J)(1 - \delta_{mm^{\prime}}) \ ,
\label{amm}
\end{eqnarray}
\begin{eqnarray} 
B^{\alpha}_{imm^{\prime}} = J (1 - \delta_{mm^{\prime}}) \ , 
\ \ \ \ \ \ \ (\alpha = x,y) \ ,
\label{bamm}
\end{eqnarray}
\begin{eqnarray} 
B^{z}_{imm^{\prime}} 
=  U_{0} \delta_{mm^{\prime}} + J (1 - \delta_{mm^{\prime}}) \ .
\label{bzmm}
\end{eqnarray}
Equation (\ref{z0}) is a partition function for a time-dependent
Hamiltonian $H(\tau, \boldsymbol{\xi}, -i\zeta)$ of an independent 
particle system.  Note that we have introduced a magnetic field 
$h^{\alpha}_{im}$ for convenience.

In the Matsubara frequency representation, the free energy ${\cal F}$ is
written as
\begin{eqnarray}
e^{-\beta {\cal F}} = 
\int \Big[ \prod_{j=1}^{N} \prod_{m=1}^{2l+1} 
\delta \boldsymbol{\xi}_{jm} \delta \zeta_{jm} \Big]
\exp \left[ - \beta E[\boldsymbol{\xi},\zeta] \,\right] \ ,
\label{free1}
\end{eqnarray}
\begin{eqnarray}
E[\boldsymbol{\xi},\zeta] & = & - \beta^{-1} \ln {\rm Tr}
({\rm e}^{-\beta H_{0}}) 
- \beta^{-1} {\rm Sp} \ln (1-vg)   \nonumber \\ 
& & \hspace{-3mm}
+ \frac{1}{4} \sum_{in} \sum_{mm^{\prime}} \bigg[ 
\zeta_{im}^{\ast}(i\omega_{n}) A_{imm^{\prime}} 
\zeta_{m^{\prime}}(i\omega_{n})   \nonumber \\
& & \hspace{-3mm}
+ \sum_{\alpha} \xi^{\ast}_{im\alpha}(i\omega_{n})
B^{\alpha}_{imm^{\prime}} \xi_{im^{\prime}\alpha}(i\omega_{n})
\bigg] \ ,  \hspace{5mm}
\label{exiz}
\end{eqnarray}
\begin{eqnarray}
(v)_{iLn\sigma jL^{\prime}n^{\prime}\sigma^{\prime}} 
= v_{jL\sigma\sigma^{\prime}}(i\omega_{n}-i\omega_{n^{\prime}})
\delta_{ij}\delta_{LL^{\prime}} \ ,
\label{dpot}
\end{eqnarray}
\begin{eqnarray}
v_{iL\sigma\sigma^{\prime}}(i\omega_{n}) & = & 
- \frac{1}{2}  \sum_{m^{\prime}} i A_{imm^{\prime}}
\zeta_{im^{\prime}}(i\omega_{n}) \delta_{l2}\delta_{\sigma\sigma^{\prime}} 
\nonumber \\ 
& &  \hspace{-25mm}
- \sum_{\alpha} \left( \frac{1}{2} \sum_{m^{\prime}} 
B^{\alpha}_{imm^{\prime}} \xi_{im^{\prime}\alpha}(i\omega_{n})
+ h^{\alpha}_{im} \right)\delta_{l2} 
(\sigma_{\alpha})_{\sigma\sigma^{\prime}} . \hspace{7mm}
\label{dpot2}
\end{eqnarray}
The functional integrals in the Fourier representation in 
eq. (\ref{free1}) is given by 
\begin{eqnarray} 
\int \Big[ \prod_{m=1}^{2l+1} \delta \zeta_{im} \Big]  \hspace{-2mm} 
& = &  \hspace{-3mm} \int \prod_{m=1}^{N^{\prime}} 
\sqrt{\dfrac{\beta^{2l+1} {\rm det} A_{i}}{(4\pi)^{2l+1}}} 
\prod_{m=1}^{2l+1} d\zeta_{im}(0)  \nonumber \\
& &    \hspace{-5mm} \times
\left[ \prod_{n=1}^{\infty}
\dfrac{\beta^{2l+1} {\rm det} A_{i}}{(4\pi)^{2l+1}}
d^{2}\zeta_{im}(i\omega_{n}) \right] .
\label{func2}
\end{eqnarray}
Here the field variable $\zeta_{im}(i\omega_{n})$ 
($\xi_{im\alpha}(i\omega_{n})$) denotes the $n$-frequency component of 
$\zeta_{im}(\tau)$ ($\xi_{im\alpha}(\tau)$), and $d^{2}\zeta_{im}(i\omega_{n})=d {\rm Re}\zeta_{im}(i\omega_{n})
d {\rm Im}\zeta_{im}(i\omega_{n})$.
The energy functional $E[\boldsymbol{\xi},\zeta]$ in eq. (\ref{exiz})
consists of the noninteracting term (the first term at the r.h.s. (the
right-hand-side)), the scattering term due to dynamical potential
(the second term), 
and the Gaussian term (the third term).  
Sp in the second term at the r.h.s. of eq. (\ref{exiz}) means a trace
over site, orbital, frequency, and spin.  $g$ in the second term
denotes the temperature Green function for noninteracting system
$H_{0}$.  The dynamical potential $v$ is defined by eqs. (\ref{dpot})
and (\ref{dpot2}), and $\sigma_{\alpha}$ in eq. (\ref{dpot2}) denotes 
the $\alpha$ component of the Pauli spin matrices.  

In the effective medium approach~\cite{kake02}, 
we introduce a coherent potential
\begin{eqnarray} 
(\Sigma)_{iLn\sigma jL^{\prime}n^{\prime}\sigma^{\prime}} =
\Sigma_{L\sigma}(i\omega_{n})
\delta_{ij}\delta_{LL^{\prime}}\delta_{nn^{\prime}}
\delta_{\sigma\sigma^{\prime}} \ ,
\label{sigma}
\end{eqnarray}
into the energy functional $E[\boldsymbol{\xi},\zeta]$, and expand it
with respect to $v-\Sigma$ as
\begin{eqnarray} 
E[\boldsymbol{\xi},\zeta] = \tilde{\cal F} 
+ \sum_{i} E_{i}[\boldsymbol{\xi}_{i},\zeta_{i}] + \Delta E \ .
\label{exiz2}
\end{eqnarray}
Here the zero-th order term $\tilde{\cal F}$ is a coherent part of the
free energy which is defined by
\begin{eqnarray} 
\tilde{\cal F}  = - \beta^{-1} {\rm ln} {\rm Tr}({\rm e}^{-\beta H_{0}}) 
- \beta^{-1} {\rm Sp} \ln (1 - \Sigma g) \ . 
\label{fcoh}
\end{eqnarray}
Note that the coherent part does not depend on the dynamical potential.

The next term in eq. (\ref{exiz2}) consists of a sum of the single-site
energies $E_{i}[\boldsymbol{\xi}_{i},\zeta_{i}]$, which are defined by
\begin{eqnarray}
E_{i}[\boldsymbol{\xi}_{i},\zeta_{i}] & = & 
- \beta^{-1} {\rm tr}\ {\rm ln}
(1 - \delta v_{i} F_{i}\,)    \nonumber \\
& & \!\!\!\!\!\!\!\!
+ \frac{1}{4} \sum_{n} \sum_{mm^{\prime}}
[\zeta^{\ast}_{im}(i\omega_{n})A_{imm^{\prime}} 
\zeta_{im^{\prime}}(i\omega_{n})   \nonumber \\
& &   \hspace{-5mm}
+ \sum_{\alpha}^{xyz} \xi^{\ast}_{im\alpha}(i\omega_{n})
B^{\alpha}_{imm^{\prime}} \xi_{im^{\prime}\alpha}(i\omega_{n}) ] \ .
\label{exizi}
\end{eqnarray}
Here tr means a trace over orbital, frequency, and spin on site $i$.
$\delta v_{i}=v_{i}-\Sigma_{i}$, and $v_{i}$ ($\Sigma_{i}$) is the
dynamical (coherent) potential on site $i$.
$F_{i}$ is the site-diagonal component of the coherent Green function 
defined by
\begin{eqnarray}
(F_{i})_{jLn\sigma j^{\prime}L^{\prime}n^{\prime}\sigma^{\prime}} =
F_{iL\sigma}(i\omega_{n})
\delta_{ij}\delta_{ij^{\prime}}\delta_{LL^{\prime}}
\delta_{nn^{\prime}}\delta_{\sigma\sigma^{\prime}} \ ,
\label{cohfi}
\end{eqnarray}
\begin{eqnarray}
F_{iL\sigma}(i\omega_{n}) = 
[(g^{-1} - \Sigma)^{-1}]_{iLn\sigma iLn\sigma} \ .
\label{cohfi2}
\end{eqnarray}

The last term in eq. (\ref{exiz2}) denotes the higher order terms in
expansion.
\begin{eqnarray}
\Delta E = - \beta^{-1} \, {\rm Sp} \ln (1-\tilde{t}F^{\prime}) \ .
\label{dlte}
\end{eqnarray}
Here $\tilde{t}$ is the single-site $t$-matrix defined by 
\begin{eqnarray}
\tilde{t} = (1 - \delta v_{i} F_{i}\,)^{-1} \delta v_{i} \ ,
\label{tmatrix}
\end{eqnarray}
and $F^{\prime}$ is the off-diagonal coherent Green Function defined by
\begin{eqnarray}
(F^{\prime})_{iLn\sigma jL^{\prime}n^{\prime}\sigma^{\prime}} \!\! = 
[(g^{-1} \!\!\! - \!\! \Sigma)^{-1}]_{iLn\sigma 
jL^{\prime}\!n^{\prime}\!\sigma^{\prime}}
(1\!\!-\!\!\delta_{ij}) \delta_{\sigma\sigma^{\prime}} .
\label{cohfij}
\end{eqnarray}

The dynamical CPA is a single-site approximation which neglects the
intersite dynamical correlations $\Delta E$.  The free energy is then
written as 
\begin{eqnarray}
{\cal F}_{\rm CPA} = \,\tilde{\!\!\cal F} 
- \!\! \sum_{i} \beta^{-1} \,{\rm ln} \!\int \![\prod_{m} \!
\delta \boldsymbol{\xi}_{im} \delta \zeta_{im}] 
{\rm e}^{\displaystyle -\beta E_{i}[\boldsymbol{\xi}_{i},\zeta_{i}]} .  
\label{fcpa}
\end{eqnarray}
The dynamical coherent potential $\Sigma_{iL\sigma}(i\omega_{n})$ 
should be determined so that the nonlocal corrections 
$\Delta E$ vanish in average.  This means that
\begin{eqnarray}
\langle \tilde{t}_{i} \rangle = 0 \ ,
\label{avt}
\end{eqnarray}
where
\begin{eqnarray}
\langle (\sim) \rangle = 
\dfrac{\displaystyle 
\int \,[\prod_{m} 
\delta\boldsymbol{\xi}_{im} \delta\zeta_{im}] (\sim)
\ {\rm e}^{\displaystyle -\beta E_{i}[\boldsymbol{\xi}_{i},\zeta_{i}]} 
}
{\displaystyle \int \,[\prod_{m} 
\delta\boldsymbol{\xi}_{im} \delta\zeta_{im}] 
\ {\rm e}^{\displaystyle -\beta E_{i}[\boldsymbol{\xi}_{i},\zeta_{i}]}
} \ .
\label{av}
\end{eqnarray}

The above condition called the CPA equation is written as 
\begin{eqnarray}
\langle G^{(i)}_{iL\sigma}(i\omega_{n}) 
\rangle = F_{iL\sigma}(i\omega_{n}) \ ,
\label{dcpa}
\end{eqnarray}
\begin{eqnarray}
G^{(i)}_{iL\sigma}(i\omega_{n}) = 
[(F_{i}^{-1} - \delta v_{i})^{-1}]_{iLn\sigma iLn\sigma} \ .
\label{igreen}
\end{eqnarray}
Here the l.h.s. (left-hand-side) of eq. (\ref{dcpa}) is 
a temperature Green function 
for an impurity system in the effective medium, whose Hamiltonian is
given as follows.
\begin{eqnarray}
H^{(i)}(\tau) & = & \tilde{H}(\tau) + H^{(i)}_{1}(\tau)  \nonumber \\
& &  \hspace{-10mm}
- \int_{0}^{\beta} d\tau^{\prime} \sum_{L\sigma} 
a_{iL \sigma}^{\dagger}(\tau) \Sigma_{iL\sigma}(\tau - \tau^{\prime})
a_{jL \sigma}(\tau^{\prime}) , \hspace{5mm}
\label{himp}
\end{eqnarray}
\begin{eqnarray}
\tilde{H}(\tau) & = & \sum_{iL\sigma} (\epsilon^{0}_{iL} - \mu) \, 
\hat{n}_{iL \sigma}(\tau)   \nonumber \\
& & \hspace{-5mm}
+ \sum_{iL jL^{\prime} \sigma} 
t_{iL jL^{\prime}} \, 
a_{iL \sigma}^{\dagger}(\tau) a_{jL^{\prime} \sigma}(\tau) 
\nonumber \\
& &  \hspace{-10mm}
+ \sum_{jL\sigma} \int_{0}^{\beta} d\tau^{\prime} 
a_{jL \sigma}^{\dagger}(\tau) \Sigma_{jL\sigma}(\tau - \tau^{\prime})
a_{jL \sigma}(\tau^{\prime}) ,  \hspace{5mm}
\label{htilde}
\end{eqnarray}
\begin{eqnarray}
H^{(i)}_{1}(\tau) & = & 
\sum_{m} U_{0} \, \hat{n}_{ilm \uparrow}(\tau) 
\hat{n}_{ilm \downarrow}(\tau)   \nonumber \\
& &
+ {\sum_{m > m^{\prime}}} 
(U_{1}-\frac{1}{2}J) \, \hat{n}_{ilm}(\tau) \hat{n}_{ilm^{\prime}}(\tau) 
\hspace{10mm}  \nonumber \\
& &
- {\sum_{m > m^{\prime}}} J \,   
\hat{\boldsymbol{s}}_{ilm}(\tau) \cdot 
\hat{\boldsymbol{s}}_{ilm^{\prime}}(\tau) \ . 
\label{h1imp}
\end{eqnarray}
It should be noted that the CPA equation (\ref{dcpa}) is equivalent to
the following stationary condition.
\begin{eqnarray}
\frac{\mathstrut \delta \mathcal{F}_{\rm CPA}}
{\mathstrut \delta\Sigma_{iL\sigma}(i\omega_{n})}
 = 0 \ .
\label{dcpa2}
\end{eqnarray}
\vspace{0mm}

\section{Harmonic Approximation to the Dynamical CPA}
\vspace{0mm}

We can rewrite the free energy (\ref{fcpa}) by means of an effective
potential projected onto the zero frequency variables 
$\boldsymbol{\xi}_{im}=\boldsymbol{\xi}_{im}(0)$ and 
$\zeta_{im}=\zeta_{im}(0)$.
\begin{eqnarray}
{\mathcal F}_{\rm CPA} & = & \tilde{\mathcal F}   \nonumber \\
& &  \!\!\!\!\!\!\!\!\!\!\!\!\!\!\!\!\!\!\!\!\!\!
- \beta^{-1} {\rm ln} \int \left[ \prod_{\alpha} 
\sqrt{\dfrac{\beta^{2l+1} {\rm det} B^{\alpha}}{(4\pi)^{2l+1}}} 
\prod_{m} d\xi_{m\alpha} \right]    \nonumber \\
& & \ \ \times
\sqrt{\dfrac{\beta^{2l+1} {\rm det} A}{(4\pi)^{2l+1}}}
\left[ \prod_{m} d\zeta_{m} \right]
\, {\rm e}^{\displaystyle -\beta E(\boldsymbol{\xi}, \zeta)} . 
\hspace{7mm}
\label{fcpa2}
\end{eqnarray}
Note that we have redefined ${\mathcal F}_{\rm CPA}$ and ${\mathcal F}$
by those per site, assuming that all the sites are equivalent to each
other.  Furthermore we omit here and in the following all the site
indices for simplicity.

The effective potential $E(\boldsymbol{\xi}, \zeta)$ in
eq. (\ref{fcpa2}) consists of the static part 
$E_{\rm st}(\boldsymbol{\xi}, \zeta)$ and the dynamical one 
$E_{\rm dyn}(\boldsymbol{\xi}, \zeta)$.
\begin{eqnarray}
E(\boldsymbol{\xi}, \zeta) 
= E_{\rm st\,}(\boldsymbol{\xi}, \zeta) 
+ E_{\rm dyn\,}(\boldsymbol{\xi}, \zeta) \ ,
\label{eeff}
\end{eqnarray}
\begin{eqnarray}
E_{\rm st\,}(\boldsymbol{\xi}, \zeta) & = & - \beta^{-1} {\rm tr}\ {\rm ln}
[1 - \delta v_{0} F_{i}\,]   \nonumber \\
& &  \hspace{-12mm}
+ \frac{1}{4} \sum_{mm^{\prime}} \,
[\zeta_{m} A_{mm^{\prime}} \zeta_{m^{\prime}}
+ \sum_{\alpha}^{xyz} \xi_{m\alpha}
B^{\alpha}_{mm^{\prime}} \xi_{im^{\prime}\alpha} ] ,  \hspace{7mm}
\label{est}
\end{eqnarray}
\begin{eqnarray}
{\rm e}^{\displaystyle -\beta E_{\rm dyn\,}(\boldsymbol{\xi},\zeta)}
& = & \overline{D}    \nonumber \\
& \hspace*{-54mm} \equiv & \hspace*{-32mm} \int \!\!\! \prod_{n=1}^{\infty}
\!\! \left[ \prod_{\alpha} \!\!\dfrac{\beta^{2l+1} \!{\rm det} 
B^{\alpha}\!\!}{(2\pi)^{2l+1}}
d^{2}\!\xi_{m\alpha}\!(i\omega_{n}) \!\right] 
\!\!\dfrac{\beta^{2l+1} \!{\rm det} A}{(2\pi)^{2l+1}}
\!\!\left[ \!\prod_{m} \!\!d^{2}\!\zeta_{m}\!(i\omega_{n}) \!\right]
\nonumber \\
&  &  \hspace*{-30mm}
\times D \, \exp \bigg[ - \frac{\beta}{4} \sum_{n \ne 0} 
\sum_{mm^{\prime}} \Big( 
\zeta_{m}^{\ast}(i\omega_{n}) A_{mm^{\prime}} 
\zeta_{m^{\prime}}(i\omega_{n})    \nonumber \\
& & \hspace{-8mm}
+ \sum_{\alpha} \xi^{\ast}_{m\alpha}(i\omega_{n})
B^{\alpha}_{mm^{\prime}} \xi_{m^{\prime}\alpha}(i\omega_{n})
\Big) \bigg] ,
\label{edyn}
\end{eqnarray}
\begin{eqnarray}
D \!\!\!& = & \!\!\!\!{\rm det} \Big(
\delta_{nn^{\prime}}\delta_{LL^{\prime}}\delta_{\sigma\sigma^{\prime}}
\nonumber \\
& &
- \sum_{\sigma^{\prime\prime}}\tilde{v}_{L\sigma\sigma^{\prime\prime}}
(i\omega_{n} - i\omega_{n^{\prime}})
\tilde{g}_{L\sigma^{\prime\prime}L^{\prime}\sigma^{\prime}}
(i\omega_{n^{\prime}})
\Big) \ .   \hspace{5mm}
\label{det}
\end{eqnarray}
Here $\delta v_{0}$ in eq. (\ref{est}) is defined by 
$\delta v_{0} = v(0) - \Sigma$ :
\begin{eqnarray}
(\delta v_{0})_{Ln\sigma L^{\prime}n^{\prime}\sigma^{\prime}} = 
(v_{L\sigma\sigma^{\prime}}(0) - 
\Sigma_{L\sigma}(i\omega_{n})\delta_{\sigma\sigma^{\prime}})
\delta_{LL^{\prime}}\delta_{nn^{\prime}} \ .
\label{dv0}
\end{eqnarray}
$v_{L\sigma\sigma^{\prime}}(0)$ is the static potential, while 
$\tilde{v}$ in eq. (\ref{det}) is the dynamical potential without
zero frequency part.
\begin{eqnarray}
\tilde{v}_{L\sigma\sigma^{\prime}}(i\omega_{n} - i\omega_{n^{\prime}}) 
= v_{L\sigma\sigma^{\prime}}
(i\omega_{n} - i\omega_{n^{\prime}}) - v_{L\sigma\sigma^{\prime}}(0)
\delta_{nn^{\prime}} \ .
\label{vdyn}
\end{eqnarray}
Furthermore,
$\tilde{g}_{L\sigma L^{\prime}\sigma^{\prime}}(i\omega_{n})$ in
eq. (\ref{det}) is the Green function in the static approximation
defined by
\begin{eqnarray}
\tilde{g}_{L\sigma L^{\prime}\sigma^{\prime}}(i\omega_{n}) = 
[(F^{-1} - \delta v_{0})^{-1}]_{Ln\sigma 
L^{\prime}n\sigma^{\prime}} \ ,
\label{gst}
\end{eqnarray}
where the coherent Green function $F$ is defined by eqs. (\ref{cohfi})
and (\ref{cohfi2}).

In the functional integral approach, we first have to
calculate the determinant (\ref{det}), and second have to evaluate the
functional integral in eq. (\ref{edyn}).  In order to implement these
calculations, we expand the determinant (\ref{det}) with respect to the
frequency modes of the dynamical potential $v_{L\sigma\sigma^{\prime}}$ 
as follows.
\begin{eqnarray}
D & = & 1 + \sum_{\nu} \,(D_{\nu} -1)   \nonumber \\
& &
+ \sum_{(\nu,\nu^{\prime})} \,(D_{\nu\nu^{\prime}} 
- D_{\nu} - D_{\nu^{\prime}} + 1) + \cdots ,   \hspace{8mm}
\label{detexp}
\end{eqnarray}
\begin{eqnarray}
D_{\nu} \!\!\!\!& = & \!\!\!\!{\rm det} \, \Big[ \,\delta_{LL^{\prime}}
\delta_{\sigma\sigma^{\prime}}\delta_{nn^{\prime}}
- \sum_{\sigma^{\prime\prime}} ( 
v_{L\sigma\sigma^{\prime\prime}}(i\omega_{\nu})\delta_{n-n^{\prime},\,\nu} 
\nonumber \\
& & + 
v_{L\sigma\sigma^{\prime\prime}}(i\omega_{-\nu})\delta_{n-n^{\prime},-\nu} ) 
\tilde{g}_{L\sigma^{\prime\prime} L^{\prime}\sigma^{\prime}}
(i\omega_{n^{\prime}}) \, \Big] ,   \hspace{5mm}
\label{detnu}
\end{eqnarray}
\begin{eqnarray}
D_{\nu\nu^{\prime}} \!\!\!\! & = & \!\!\!\! 
{\rm det} \, \Big[ \,\delta_{LL^{\prime}}
\delta_{\sigma\sigma^{\prime}}\delta_{nn^{\prime}} \nonumber \\
& &  \hspace{-16mm}
- \!\!\sum_{\sigma^{\prime\prime}} ( 
v_{L\sigma\!\sigma^{\prime\prime}}(i\omega_{\nu}\!)
\delta_{n\!-\!n^{\prime}, \nu} \!\!+\!\! 
v_{L\sigma\!\sigma^{\prime\prime}}(i\omega_{\!-\!\nu}\!)
\delta_{n\!-\!n^{\prime}\!,-\nu} ) 
\tilde{g}_{L\sigma^{\prime\prime} \!L^{\prime}\!\sigma^{\prime}}
(i\omega_{n^{\prime}}\!)  \nonumber \\
& & \hspace*{-16mm}
- \!\sum_{\sigma^{\prime\prime}} ( 
v_{L\!\sigma\!\sigma^{\prime\prime}}\!(i\omega_{\nu^{\prime}}\!)
\delta_{n\!-\!n^{\prime}\!, \nu^{\prime}} \nonumber \\
& & \hspace{5mm}
 + v_{L\sigma\!\sigma^{\prime\prime}}\!(i\omega_{\!-\!\nu^{\prime}}\!)
\delta_{n\!-\!n^{\prime}\!,-\nu^{\prime}} \!) 
\tilde{g}_{L\!\sigma^{\prime\prime} \!L^{\prime}\!\sigma^{\prime}}
(i\omega_{n^{\prime}}) \, \Big] .
\label{detnunu}
\end{eqnarray}
The first term at the r.h.s. of eq. (\ref{detexp}) 
corresponds to the zero-th
approximation ({\it i.e.} the static approximation) which neglects 
dynamical potentials.  The second term is a superposition of
the independent scattering terms of dynamical potential 
$v_{L\sigma\sigma^{\prime}}(i\omega_{\nu})$.
Higher order terms describe dynamical mode-mode couplings.

We adopt here the harmonic approximation~\cite{amit71} 
which neglects the mode-mode
coupling terms in eq. (\ref{detexp}). We have then
\begin{eqnarray}
E_{\rm dyn}(\boldsymbol{\xi},\zeta) = - \beta^{-1} {\rm ln} 
\left[ 1 + \sum_{\nu} \,(\overline{D}_{\nu} -1) \right] .
\label{edyn2}
\end{eqnarray}
The approximation yields the result of the second-order perturbation in
the weak Coulomb interaction limit, and describes the Kondo
anomaly in the strong interaction limit~\cite{dai91}.

Let us now calculate $\overline{D}_{\nu}$ in eq. (\ref{edyn2}).
The determinant $D_{\nu}$ in the harmonic approximation is written by a
product of those of the tridiagonal-type matrices as
\begin{eqnarray}
D_{\nu} = \prod_{k=0}^{\nu-1} \left[ \prod_{m=1}^{2l+1} D_{\nu}(k,m)
\right] ,
\label{dnu2}
\end{eqnarray}
\begin{eqnarray}
D_{\!\nu}(\!k,\!m\!) \!\!= \!\!\left| 
\begin{array}{@{\,}ccccccc@{\,}}
\ddots\!\!\!\!\!\!\!\!\!\!\!\!\! &         &         &      &    &   &  \\
       & 1       & 1           &        & 0  & &  \\
       & a_{\!-\!\nu\!+\!k}(\!\nu,\!m\!)\!\!\!\!\!\!\!\! & 1    & 1    &    & &  \\
       &                & a_{k}\!(\!\nu,\!m)\!\!\!\!\!\!\!\!\!\!\! & 1   & 1  & &  \\
       &            &      & a_{\nu\!+\!k}(\!\nu,\!m)\!\!\!\!\!\!\!\!\!\!\! & 1  & 1 & \\
       & 0       &      &       & a_{2\nu\!+\!k}(\!\nu,\!m)\!\!\!\!\!\!\!\! & & \\
       &          &      &          &                & \!\!\!\ddots \!\!\!\!\!\!\!\!\!\!\!\!& \\
\end{array}
\right| . \hspace{-2mm}
\label{dnukm}
\end{eqnarray}
Here 1 in the determinant is the $2 \times 2$ unit matrix, 
$a_{n}(\nu,m)$ is a $2 \times 2$ matrix
defined by 
\begin{eqnarray}
a_{n}(\nu,m)_{\sigma\sigma{\prime}} \!\!\! & = & \!\!\!\!\!\!
\sum_{\sigma^{\prime\prime}\sigma^{\prime\prime\prime}
\sigma^{\prime\prime\prime\prime}} 
v_{L\sigma\sigma^{\prime\prime}}(\nu)  \nonumber \\
& & \hspace{-15mm} \times
\tilde{g}_{L\sigma^{\prime\prime} \sigma^{\prime\prime\prime}}(n-\nu)
v_{L\sigma^{\prime\prime\prime}\sigma^{\prime\prime\prime\prime}}(-\nu) 
\tilde{g}_{L\sigma^{\prime\prime\prime\prime} \sigma^{\prime}}(n) \ .
\hspace{5mm}
\label{annum}
\end{eqnarray}
We assumed in the above expression 
that the orbitals $\{ L \}$ form an irreducible
representation of the point group of the system, so that 
$\tilde{g}_{L\sigma L^{\prime}\sigma^{\prime}}(i\omega_{n})=
\tilde{g}_{L\sigma\sigma^{\prime}}(i\omega_{n})\delta_{LL^{\prime}}$ 
(see eq. (\ref{gst})).
Furthermore here and in the following, we write the frequency
dependence, for example, of 
$\tilde{g}_{L\sigma\sigma^{\prime}}(i\omega_{n})$ as
$\tilde{g}_{L\sigma\sigma^{\prime}}(n)$ for simplicity.

The determinant $D_{\nu}(k,m)$ is expanded with respect to the dynamical
potentials as follows.
\begin{eqnarray}
D_{\nu}(k,m) = 1 + D^{(1)}_{\nu}(k,m) + D^{(2)}_{\nu}(k,m) + \cdots ,
\hspace{3mm}
\label{dnukm2}
\end{eqnarray}
\begin{eqnarray}
D^{(n)}_{\nu}(k,m)  \hspace{-3mm} & = & \hspace{-5mm}
\sum_{\alpha_{1}\gamma_{1} \cdots \alpha_{n}\gamma_{n}}
v_{\alpha_{1}}(\nu,m)v_{\gamma_{1}}(-\nu,m) \cdots  \nonumber \\
& &  \hspace{-5mm} \times
v_{\alpha_{n}}(\nu,m)v_{\gamma_{n}}(-\nu,m) 
\hat{D}^{(n)}_{\{ \alpha\gamma \}}(\nu,k,m) \ .  \hspace{5mm} 
\label{dnnukm}
\end{eqnarray}
Here the subscripts 
$\alpha_{i}$ and $\gamma_{i}$ take 4 values $0$, $x$, $y$, and $z$,
and
\begin{eqnarray}
v_{0}(\nu,m) = - \dfrac{1}{2} i \sum_{m^{\prime}} A_{mm^{\prime}}
\zeta_{m^{\prime}}(\nu)\delta_{l2} \ ,
\label{v0num}
\end{eqnarray}
\begin{eqnarray}
v_{\alpha}(\nu,m) = - \dfrac{1}{2} \sum_{m^{\prime}} 
B^{\alpha}_{mm^{\prime}} \xi_{m^{\prime}\alpha}(\nu)\delta_{l2} \ , 
\hspace*{2mm} (\alpha=x,y,z) .
\label{vanum}
\end{eqnarray}
Note that the subscript 
$\{ \alpha\gamma \}$ in eq. (\ref{dnnukm}) denotes a set of 
$(\alpha_{1}\gamma_{1} \cdots \alpha_{n}\gamma_{n})$.
The expressions of $\hat{D}^{(n)}_{\{ \alpha\gamma \}}(\nu,k,m)$ are
given in Appendix A.

Substituting eq. (\ref{dnukm2}) into eq. (\ref{dnu2}) and taking the
Gaussian average (\ref{edyn}), we have
\begin{eqnarray}
\overline{D_{\nu}} \!\!\!\!\! &=& \!\!\!\!\! \sum_{n=0}^{\infty} 
\sum_{\alpha_{1}\!\gamma_{1} \cdots \alpha_{n}\!\gamma_{n}}  \!\!
\sum_{\sum_{km} \!l(\!k,\!m)=n} \!\!\!\!
\overline{\Big[ \prod_{m=1}^{2l+1} \!\! \prod_{i} \!
v_{\alpha_{i}}\!(\nu,\!m)v_{\gamma_{i}}\!(\!-\nu,\!m) \Big]}   \nonumber \\
& &  \hspace{20mm} \times
\Big[ \prod_{m=1}^{2l+1} \prod_{k=0}^{\nu-1}
\hat{D}^{(l(k,m))}_{\{ \alpha\gamma \} }(\nu,k,m) \Big] . 
\label{dnubar}
\end{eqnarray}
Here 
$\{ l(k,m)\} (k=0, \cdots, \nu-1, m=1, \cdots , 2l+1)$ are zero or
positive integer, satisfying $\sum_{km} l(k,m)=n$.
$\prod_{i} v_{\alpha_{i}}(\nu,m)v_{\gamma_{i}}(-\nu,m)$
are the products of $v_{\alpha_{i}}(\nu,m)v_{\gamma_{i}}(-\nu,m)$
belonging to the $m$-th orbital block.
Calculations of the Gaussian average of the dynamical potentials 
are given in Appendix B, and we reach the following expression. 
\begin{eqnarray}
\overline{D}_{\nu} = 1 + \overline{D}^{(1)}_{\nu} + 
\overline{D}^{(2)}_{\nu} + \cdots \ , 
\label{dnubar2}
\end{eqnarray}
\begin{eqnarray}
\overline{D}^{(n)}_{\nu}  \!\!\!\! &=& \!\!\!\! \dfrac{1}{(2\beta)^{n}} 
\!\!\! \sum_{\sum_{km} l(k,m)=n} \sum_{\{ \alpha_{j}(\!k,\!m)\} }
\sum_{\rm P}   \nonumber \\
& &  \hspace{-8mm}
\prod_{m=1}^{2l+1} \prod_{k=0}^{\nu-1}
\Bigg[ \Big( \prod_{j=1}^{l(k,m)} C^{\alpha_{j}(k,m)}_{mm_{\rm p}} \Big)
\hat{D}^{(l(k,m))}_{\{ \alpha\alpha_{{\rm p}^{-1}} \} }(\nu,k,m) \Bigg]
. \hspace{5mm}
\label{dnubarn}
\end{eqnarray}
Here $j$ denotes the $j$-th member
of the $(k,m)$ block. P denotes a permutation of a set $\{ (j,k,m) \}$: 
${\rm P} \{ (j,k,m) \} = \{ (j_{\rm p},k_{\rm p},m_{\rm p}) \}$, 
$\alpha_{{\rm p}^{-1}}$ 
means an rearrangement of $\{ \alpha_{j}(k,m) \}$ according to 
the inverse permutation P${}^{-1}$.  Note that $\alpha_{j}(k,m)$ takes 4
values $0$, $x$, $y$, and $z$.
$C^{\alpha}_{mm^{\prime}}$ is a Coulomb interaction defined by 
\begin{eqnarray}
C^{\alpha}_{mm^{\prime}} = \begin{cases}
-A_{mm^{\prime}} & (\alpha=0) \\
B^{\alpha}_{mm^{\prime}}  & (\alpha=x,y,z) \ .
\end{cases}
\label{cdef}
\end{eqnarray}
Equations (\ref{edyn2}) and (\ref{dnubar2}) determine the dynamical
potential $E_{dyn}(\boldsymbol{\xi}, \zeta)$.

The free energy (\ref{fcpa2}) is written alternatively as
\begin{eqnarray}
{\cal F}_{\rm CPA} \!\!\!\!&=& \!\!\! \tilde{\!\!\cal{F}} 
- \beta^{-1} {\rm ln}  \int \Bigg[ \prod_{\alpha} 
\sqrt{\dfrac{\beta^{2l+1} {\rm det} B^{\alpha}}{(4\pi)^{2l+1}}}
\prod_{m} 
d\boldsymbol{\xi}_{m} \Bigg]   \nonumber \\
& &  \hspace{35mm} \times
{\rm e}^{\displaystyle -\beta E_{\rm eff}(\boldsymbol{\xi})} \ . 
\hspace{10mm}
\label{fcpa3}
\end{eqnarray}
In the itinerant electron system, spin fluctuations plays an important
role, and we may neglect the thermal charge fluctuations making use of 
the saddle-point approximation to the static charge
fields $\zeta_{m}$.  We have then 
$E_{\rm eff}(\boldsymbol{\xi})=E(\boldsymbol{\xi}, 
\zeta^{\ast})$.
The saddle point value $\zeta^{\ast}_{m}$ is determined from 
$\partial E(\boldsymbol{\xi}, \zeta^{\ast}) / \partial \zeta_{m} = 0$:
\begin{eqnarray}
- i \zeta^{\ast}_{m} = \tilde{n}_{L}(\boldsymbol{\xi}) 
= \sum_{\sigma} \tilde{n}_{L\sigma}(\boldsymbol{\xi}) \ ,
\label{saddn}
\end{eqnarray}
\begin{eqnarray}
\tilde{n}_{L\sigma}(\boldsymbol{\xi}) = \dfrac{1}{\beta} 
\sum_{n} G_{L\sigma}(n) \ .
\label{saddn2}
\end{eqnarray}

In order to reduce the number of variables, we neglect the out-of-phase 
thermal spin fluctuations between different orbitals on a site, 
and take into account
their in-phase fluctuations.  This can be made by introducing a large
variable $\xi_{\alpha} = \sum_{m} \xi_{m\alpha}$.  Inserting 
$1 = \int [\prod_{\alpha} d\xi_{\alpha} d\lambda_{\alpha}] \exp [-2\pi i
\lambda_{\alpha} (\xi_{\alpha} - \sum_{m} \xi_{m\alpha})]$ into
eq. (\ref{fcpa3}), and replacing variables $\xi_{m\alpha}$ with
$\xi_{\alpha}/(2l+1)$ in the non-Gaussian terms of 
$E_{\rm eff}(\boldsymbol{\xi})$, we reach 
\begin{eqnarray}
{\cal F}_{\rm CPA} = \ \tilde{\!\!\cal{F}} 
- \beta^{-1} {\rm ln}  \int \, \Big[ \prod_{\alpha} 
\sqrt{\dfrac{\beta \tilde{J}_{\alpha}}{4\pi}}
d \xi_{\alpha} \Big]
\,{\rm e}^{\displaystyle -\beta E_{\rm eff}(\boldsymbol{\xi})} ,
\label{fcpa4}
\end{eqnarray}
\begin{eqnarray}
E_{\rm eff\,}(\boldsymbol{\xi}) 
= E_{\rm st\,}(\boldsymbol{\xi}) 
+ E_{\rm dyn\,}(\boldsymbol{\xi}) \ ,
\label{eeff2}
\end{eqnarray}
\begin{eqnarray}
E_{\rm st\,}\!(\boldsymbol{\xi}) \!\!\!\!\!\! &=& \!\!\!\!\!\! 
- \dfrac{1}{\beta} \sum_{mn} 
{\rm ln} \Big[ 
(1 \!\! - \!\! \delta v_{\!L\!\uparrow}(0)F_{\!L\!\uparrow}(n))
(1 \!\! - \!\! \delta v_{\!L\!\downarrow}(0)F_{\!L\!\downarrow}(n)) \nonumber \\
& &  \hspace{30mm}
- \dfrac{1}{4} \tilde{J}^{2}_{\bot} \xi^{2}_{\bot} 
F_{L\uparrow}(n)F_{L\downarrow}(n)
\Big]     \nonumber \\
&  & 
+ \dfrac{1}{4} \Big[
- (U_{0}-2U_{1}+J) \sum_{m} \tilde{n}_{L}(\boldsymbol{\xi})^{2}
\nonumber \\
& &  \hspace{7mm}
- (2U_{1}-J) \tilde{n}_{l}(\boldsymbol{\xi})^{2}
+ \tilde{J}^{2}_{\bot} \xi^{2}_{\bot} + \tilde{J}^{2}_{z} \xi^{2}_{z}
\Big] .
\label{est2}
\end{eqnarray}
Here $\tilde{J}_{x}=\tilde{J}_{y}=\tilde{J}_{\bot}=(1-1/(2l+1))J$, 
$\tilde{J}_{z} = U_{0}/(2l+1) + \tilde{J}_{\bot}$, 
$\delta v_{L\sigma}(0) = v_{L\sigma}(0) -
\Sigma_{L\sigma}(n)$, and 
$v_{L\sigma}(0) = v_{0}(0,m) + \sigma v_{z}(0,m)$.
The charge densities, $\tilde{n}_{L}(\boldsymbol{\xi})$ and 
$\tilde{n}_{l}(\boldsymbol{\xi})$ are defined by 
$\tilde{n}_{L}(\boldsymbol{\xi})=\sum_{\sigma}
\tilde{n}_{L\sigma}(\boldsymbol{\xi})$ and 
$\tilde{n}_{l}(\boldsymbol{\xi})=\sum_{m}
\tilde{n}_{L}(\boldsymbol{\xi})$.
Furthermore $E_{\rm dyn\,}(\boldsymbol{\xi})$ is given by 
eq. (\ref{edyn2}) in which $\zeta_{m}$  ($\xi_{m\alpha}$)
has been replaced by $i\tilde{n}_{L}(\boldsymbol{\xi})$
($\xi_{\alpha}/(2l+1)$).

The CPA equation in the HA is obtained from the stationary condition
(\ref{dcpa2}) with the free energy (\ref{fcpa4}).
\begin{eqnarray}
\langle G_{L\sigma}(n) \rangle 
= F_{L\sigma}(n) \ ,
\label{dcpa3}
\end{eqnarray}
and
\begin{eqnarray}
\langle G_{L\sigma}(n) \rangle = 
\left\langle \tilde{g}_{L\sigma\sigma}(n) - 
\dfrac{\beta}{\kappa_{L\sigma}(n)}
\dfrac{\delta E_{\rm dyn}}{\delta \Sigma_{L\sigma}(n)}
\right\rangle \ .
\label{avg}
\end{eqnarray}
Here $\kappa_{L\sigma}(n) = 1 - 
F_{L\sigma}(n)^{-2}H_{L\sigma}(n)$ and 
$H_{L\sigma}(n) = \delta F_{L\sigma}(n) / \delta \Sigma_{L\sigma}(n)$.
The average $\langle \sim \rangle$ at the r.h.s. of eq. (\ref{avg})
is now defined by a classical average with
respect to the effective potential (\ref{eeff2}).
\begin{eqnarray}
\langle \sim \rangle = \dfrac{\displaystyle \int \, \Big[ \prod_{\alpha} 
d \xi_{\alpha} \Big] (\sim)
\,{\rm e}^{\displaystyle -\beta E_{\rm eff}(\boldsymbol{\xi})}}
{\displaystyle \int \, \Big[ \prod_{\alpha} 
d \xi_{\alpha} \Big]
\,{\rm e}^{\displaystyle -\beta E_{\rm eff}(\boldsymbol{\xi})}} \ .
\label{av2}
\end{eqnarray}
Substituting eq. (\ref{edyn2}) into Eq. (\ref{avg}), we obtain 
the expression
\begin{eqnarray}
\langle G_{L\sigma}(n) \rangle = 
\left\langle \tilde{g}_{L\sigma\sigma}(n) - 
\dfrac{\displaystyle \sum_{\nu} \frac{\delta \overline{D}_{\nu}}
{\displaystyle \kappa_{L\sigma}(n)\delta \Sigma_{L\sigma}(n)}}
{1+ \sum_{\nu} (\overline{D}_{\nu}-1)}
\right\rangle \ .
\label{avg2}
\end{eqnarray}

The local charge and magnetic moment are obtained from 
$\partial {\cal F}_{\rm CPA} / \partial \epsilon^{0}_{L}$ and
$ - \partial {\cal F}_{\rm CPA} / \partial h^{\alpha}_{L}$.
Making use of the stationary conditions of ${\cal F}_{\rm CPA}$ with
respect to $\zeta^{\ast}_{m}$ and $\Sigma_{L\sigma}$, and the CPA
equation (\ref{dcpa3}), we reach
\begin{eqnarray}
\langle \hat{n}_{L} \rangle = 
\dfrac{1}{\beta} \sum_{n\sigma} F_{L\sigma}(n) \ ,
\label{avnl}
\end{eqnarray}
\begin{eqnarray}
\langle \hat{m}^{z}_{L} \rangle = 
\dfrac{1}{\beta} \sum_{n\sigma} \sigma F_{L\sigma}(n) \ .
\label{avml}
\end{eqnarray}
In particular, the $l=2$ components of local charge and 
magnetic moment are expressed as
\begin{eqnarray}
\langle \hat{n}_{l} \rangle = 
\langle \tilde{n}_{l}(\boldsymbol{\xi}) \rangle \ , 
\label{avnd}
\end{eqnarray}
\begin{eqnarray}
\langle \hat{\boldsymbol{m}}_{l} \rangle = 
\langle \boldsymbol{\xi} \rangle \ . 
\label{avmd}
\end{eqnarray}

The amplitude of charge and local moments for $d$ electrons 
are calculated from the formulae.
\begin{eqnarray}
\langle \hat{n}_{l}^{2} \rangle = \langle \hat{n}_{l} \rangle 
+ 2 \sum_{m} \dfrac{\partial {\cal F}_{\rm CPA}}{\partial U_{mm}}
+ {\sum_{mm^{\prime}}}^{\prime} \,
\dfrac{\partial {\cal F}_{\rm CPA}}{\partial U_{mm^{\prime}}} \ , 
\label{avn2l}
\end{eqnarray}
\begin{eqnarray}
\langle \boldsymbol{m}_{l}^{2} \rangle & = & 3 \langle n_{l} \rangle 
- 6 \sum_{m} \dfrac{\partial {\cal F}_{\rm CPA}}{\partial U_{mm}}
\nonumber \\
& &  \hspace{5mm}
- {\sum_{mm^{\prime}}}^{\prime} 
\left( \dfrac{\partial {\cal F}_{\rm CPA}}{\partial U_{mm^{\prime}}}
+ 2 \dfrac{\partial {\cal F}_{\rm CPA}}{\partial J_{mm^{\prime}}}
\right) \ . \hspace{5mm}
\label{avm2l}
\end{eqnarray}
Here we have introduced for convenience orbital-dependent Coulomb and
exchange interactions $U_{mm^{\prime}}$ and $J_{mm^{\prime}}$ into the
interaction $H_{1}$ to derive the expressions.
Making use of the stationary conditions of ${\cal F}_{\rm CPA}$ and
integrations by parts, we obtain
\begin{eqnarray}
\langle \hat{n}_{l}^{2} \rangle & = & 
\langle \tilde{n}_{l}(\boldsymbol{\xi}) \rangle
+ \frac{1}{2} \sum_{m} \langle \tilde{n}_{L}(\boldsymbol{\xi})^{2} \rangle 
+ {\sum_{mm^{\prime}}}^{\prime}  
\langle \tilde{n}_{L}(\boldsymbol{\xi}) 
\tilde{n}_{L^{\prime}}(\boldsymbol{\xi}) \rangle   \nonumber \\
& &
- \dfrac{1}{2(2l+1)} \left( \langle \xi^{2}_{z} \rangle - 
\dfrac{2}{\beta \tilde{J}_{z}} \right)   \nonumber \\
&  &  \hspace{-10mm}
+ \, 2 \sum_{m} \left\langle \left[ \dfrac{\partial E_{\rm dyn}}
{\partial U_{mm}} \right]_{v} \right\rangle
+ {\sum_{mm^{\prime}}}^{\prime} 
\left\langle \left[ \dfrac{\partial E_{\rm dyn}}{\partial U_{mm}}
\right]_{v} \right\rangle \ ,
\label{avn2l2}
\end{eqnarray}
\begin{eqnarray}
\langle \hat{\boldsymbol{m}}_{l}^{2} \rangle \!\!\!\!\! & = & \!\!\!\!\!
3 \langle \tilde{n}_{l}(\boldsymbol{\xi}) \rangle
- \frac{3}{2} \sum_{m} \langle \tilde{n}_{L}(\boldsymbol{\xi})^{2}
\rangle 
\nonumber \\
& &
+ \dfrac{3}{2(2l+1)} \left( \langle \xi^{2}_{z} \rangle - 
\dfrac{2}{\beta \tilde{J}_{z}} \right)  \nonumber \\
&  & \hspace{-10mm}
+ \left( 1 - \frac{1}{2l+1} \right) 
\sum_{\alpha=x,y} \left( \langle \xi^{2}_{\alpha} \rangle - 
\dfrac{2}{\beta \tilde{J}_{\alpha}} \right)   \nonumber \\
&  & \hspace{-15mm}
- 6 \sum_{m} \left\langle \left[ \dfrac{\partial E_{\rm dyn}}
{\partial U_{mm}} \right]_{v} \right\rangle
- {\sum_{mm^{\prime}}}^{\prime} 
\bigg( \left\langle \left[ \dfrac{\partial E_{\rm dyn}}
{\partial U_{mm^{\prime}}} \right]_{v} \right\rangle \nonumber \\
& &  \hspace{25mm} +
2 \left\langle \left[ \dfrac{\partial E_{\rm dyn}}
{\partial J_{mm^{\prime}}} \right]_{v} \right\rangle 
\bigg) \ .   \hspace{5mm}
\label{avm2l2}
\end{eqnarray}
Here $\left[ \partial E_{\rm dyn} / \partial U_{mm^{\prime}}
\right]_{v}$ 
means taking derivative of $E_{\rm dyn}$ with respect to 
$U_{mm^{\prime}}$ fixing the static potentials
$v_{L\sigma\sigma^{\prime}}(0)$.
In the HA, these values are obtained from eq. (\ref{edyn2}) as
\begin{eqnarray}
\left[ \dfrac{\partial E_{\rm dyn}}
{\partial U_{mm^{\prime}}} \right]_{v} = 
- \frac{1}{\beta} \, \dfrac{\displaystyle \sum_{\nu=1}^{\infty} \left[ 
\dfrac{\partial \overline{D}_{\nu}}
{\partial U_{mm^{\prime}}} \right]_{v}
}{\displaystyle 1+ \sum_{\nu=1}^{\infty} (\overline{D}_{\nu}-1)} \ ,
\label{dedyndu}
\end{eqnarray}
\begin{eqnarray}
\left[ \dfrac{\partial E_{\rm dyn}}{\partial J_{mm^{\prime}}} \right]_{v} = 
- \frac{1}{\beta} \, \dfrac{\displaystyle 
\sum_{\nu=1}^{\infty} \left[ 
\dfrac{\partial \overline{D}_{\nu}}{\partial J_{mm^{\prime}}} \right]_{v}
}{\displaystyle 1+ \sum_{\nu=1}^{\infty} (\overline{D}_{\nu}-1)} \ .
\label{dedyndj}
\end{eqnarray}

The entropy is calculated from 
$\beta^{2} \partial {\cal F}_{\rm CPA} / \partial \beta$ as
\begin{eqnarray}
S & = & \beta^{2} \dfrac{\partial \tilde{\cal F}}{\partial \beta}
+ \left\langle \beta^{2} \dfrac{\partial E_{\rm eff}}{\partial \beta} 
\right\rangle   \nonumber \\
& & \hspace{-13mm}
+ {\rm ln} \!\! \int \!\! \Big[ \!\prod_{\alpha} \!
\sqrt{\dfrac{\beta \tilde{J}_{\alpha}}{4\pi}}
d \xi_{\alpha} \! \Big]
{\rm e}^{\displaystyle -\!\beta (E_{\rm eff}(\boldsymbol{\xi}) \!
-\! \langle E_{\rm eff}(\boldsymbol{\xi}) \!\rangle )} \!-\!\dfrac{3}{2} .
\label{entropy}
\end{eqnarray}
Here
\begin{eqnarray}
\beta^{2} \dfrac{\partial \tilde{\cal F}}{\partial \beta}
= \dfrac{1}{N} {\rm Sp} \, {\rm ln} \, (g^{-1} - \Sigma)
+ \sum_{n} \sum_{L\sigma} F_{L\sigma}(n) \ ,
\label{dfcohdb}
\end{eqnarray}
\begin{eqnarray}
\left\langle \beta^{2} \dfrac{\partial E_{\rm eff}}{\partial \beta} 
\right\rangle & = & \langle \, {\rm tr} \ln \, (1 - \delta v_{0} F) \, \rangle 
- \beta \langle E_{\rm dyn} \rangle  \nonumber \\
& &  \hspace{10mm}
+ \left\langle \beta \left[ \dfrac{\partial (\beta E_{\rm dyn})}
{\partial \beta} \right]_{\omega\Sigma}
\right\rangle . \hspace{5mm}
\label{dedb}
\end{eqnarray}

The first term at the r.h.s. of eq. (\ref{entropy}) 
({\it i.e.} eq. (\ref{dfcohdb})) is the contribution from the coherent
free energy and reduces to the entropy $S_{0}$ for noninteracting
electrons when $\Sigma_{L\sigma} \longrightarrow 0$:
\begin{eqnarray}
S_{0} & = & -2 \int d\omega \rho^{0}(\omega) \big[
f(\omega) \ln f(\omega)  \nonumber \\
& &  \hspace{10mm}
+ (1-f(\omega)) \ln (1-f(\omega))
\big] \ , \hspace{5mm}
\label{s0}
\end{eqnarray}
where $\rho^{0}(\omega)$ is the total density of states per spin for
noninteracting electrons, $f(\omega)$ is the Fermi distribution
function. 
The second term in eq. (\ref{entropy}) ({\it i.e.} eq. (\ref{dedb})) is
the entropy due to the temperature dependence of the effective
potential.
$\left[ \partial (\beta E_{\rm dyn}) /
\partial \beta \right]_{\omega\Sigma}$ in eq. (\ref{dedb}) means to take
the derivative with respect to $\beta$ 
fixing the frequency $i\omega_{n}$ and the coherent
potential $\Sigma_{L\sigma}(i\omega_{n})$.  It is given in the HA as 
\begin{eqnarray}
\beta \left[ \dfrac{\partial (\beta E_{\rm dyn})}
{\partial \beta} \right]_{\omega\Sigma} = 
\dfrac{\displaystyle \sum_{\nu=1}^{\infty} \sum_{n=1}^{\infty} 
n \overline{D}^{(n)}_{\nu}}
{\displaystyle 1+ \sum_{\nu=1}^{\infty} (\overline{D}_{\nu}-1)} \ .
\label{dedb2}
\end{eqnarray}
The third and fourth terms in eq. (\ref{entropy}) produce the magnetic
entropy due to thermal spin fluctuations.

The thermodynamic energy is obtained from the relation 
$\langle H - \mu N \rangle = {\cal F}_{\rm CPA} + \beta^{-1} S$ as 
\begin{eqnarray}
\langle H - \mu N \rangle & = & \frac{1}{\beta} \sum_{n} 
\sum_{L\sigma} i\omega_{n} F_{L\sigma}(n)    \nonumber \\
&  &   \hspace{-20mm}
- \dfrac{1}{4} \bigg[
(U_{0}-2U_{1}+J) \sum_{m} 
\langle \tilde{n}_{L}(\boldsymbol{\xi})^{2} \rangle \nonumber \\
& &   \hspace{-25mm}
+ (2U_{1}-J) \langle \tilde{n}_{l}(\boldsymbol{\xi})^{2} \rangle
- \sum_{\alpha} \tilde{J}_{\alpha} 
\left( \langle \xi^{2}_{\alpha} \rangle 
- \dfrac{2}{\beta\tilde{J}_{\alpha}} \right) \bigg]    \nonumber \\
&  &   \hspace{10mm}
+ \beta^{-1} \left\langle 
\dfrac{\displaystyle \sum_{\nu=1}^{\infty} \sum_{n=1}^{\infty} 
n \overline{D}^{(n)}_{\nu}}{\displaystyle 
1+ \sum_{\nu=1}^{\infty} (\overline{D}_{\nu}-1)} 
\right\rangle .  \hspace{5mm}
\label{avenergy}
\end{eqnarray}
The first term at the r.h.s. of eq. (\ref{avenergy}) is the coherent
contribution of the kinetic energy, the second term corresponds to the
double counting correction in the Hartree-Fock energy.  The last one is
the dynamical correction to the energy.

The sum rule $n_{0}=\sum_{L} \langle \hat{n}_{L} \rangle$ determines the
chemical potential for a given valence electron number $n_{0}$.
The CPA equation (\ref{dcpa3}) and effective potential (\ref{eeff2}) with 
eqs. (\ref{est2}), (\ref{edyn2}), and (\ref{dnubar2}) form the
self-consistent equations to determine the dynamical coherent potential 
$\{ \Sigma_{L\sigma}(i\omega_{n})\}$.  After having solved the
self-consistent equations, we can calculate the magnetic moments and
charge from eqs. (\ref{avnl}), (\ref{avml}), and (\ref{avmd}), the
square of local charge and spin fluctuations from eqs. (\ref{avn2l2})
and (\ref{avm2l2}), as well as the other thermodynamic quantities 
(see eqs. (\ref{fcpa4}), (\ref{entropy}), and (\ref{avenergy})).
\vspace{10mm}

\section{Numerical calculations: Fe and Ni}   
The simplest approximation to the dynamical CPA is to neglect the
dynamical potential $E_{\rm dyn\,}(\boldsymbol{\xi})$ in the
self-consistent equations. 
This is called the static approximation and
may be justified in the high temperature limit.
The next approximation is to add the dynamical potential 
$E_{\rm dyn\,}(\boldsymbol{\xi})$ by taking into account the 
higher-order terms $\overline{D}^{(n)}$ ($n \ge 1$) 
in a series expansion (\ref{dnubar2}).
We have taken into account the terms up to the second 
order ($n \le 2$) in eq. (\ref{dnubar2}).  We call this level of
approximation the second-order dynamical CPA.  Within the approximation, 
we have performed numerical calculations for Fe and Ni in order to
examine the quantitative aspects of the theory and 
the dynamical effects on their electronic and magnetic properties.

We obtained the intraorbital
Coulomb interaction $U_{0}$, interorbital Coulomb interaction $U_{1}$, 
and exchange interaction energy parameter $J$ from the 
parameters $\overline{U}$ and $\overline{J}$ in the LDA + U via 
the relations: 
$U_{0} = \overline{U} + 8\overline{J}/5$, 
$U_{1} = \overline{U}-2\overline{J}/5$ and $J=\overline{J}$. 
(Note that $U_{0} = U_{1}+2J$.)  
We adopted in the present calculations the LDA+U values used by Anisimov
{\it et. al.}~\cite{anisimov97} ; 
$\overline{U}= 0.1691$ Ry and $\overline{J}=0.0662$ Ry 
for Fe, and
$\overline{U}=0.2205$ Ry and $\overline{J}=0.0662$ Ry for Ni.  
These values yield 
$U_{0} = 0.2749$ Ry, $U_{1} = 0.1426$ Ry, $J=0.0662$ Ry for Fe, 
and $U_{0}= 0.3263$ Ry,
$U_{1}=0.1940$ Ry, and $J=0.0662$ Ry for Ni, respectively.  

In the numerical calculations,
we adopted an approximate expression of the coherent Green
function~\cite{kirk70}  
\begin{eqnarray}
F_{L\sigma}(n) = \int \dfrac{\rho_{L}(\epsilon) d \epsilon}
{i\omega_{n} - \epsilon - \Sigma_{L\sigma}(i\omega_{n})} \ .
\label{cohg2}
\end{eqnarray}
The expression takes into account the effect of hybridization between
different $l$ blocks in the nonmagnetic state via the local densities of
states $\rho_{L}(\epsilon)$, but neglects that in the spin polarized
state. Moreover, we adopted a decoupling approximation~\cite{hase79} 
to the thermal
average of the impurity Green function in the dynamical CPA equation 
(\ref{dcpa3}).
\begin{eqnarray}
\langle G_{L\sigma}(n, \xi_{z}, \xi^{2}_{\perp}) \rangle & = &
\sum_{q=\pm} \frac{1}{2}
\left( 1 + q \dfrac{\langle \xi_{z} \rangle}
{\sqrt{\langle \xi^{2}_{z} \rangle}} \right) \nonumber \\
& &  \hspace{5mm} \times
G_{L\sigma}(n, q\sqrt{\langle \xi^{2}_{z} \rangle}, 
\langle \xi^{2}_{\perp} \rangle) \ .  \hspace{5mm}
\label{gapprox}
\end{eqnarray}
The approximation is correct up to the second moment and reasonably
describe the thermal spin fluctuations.
\begin{figure}
\includegraphics[width=9cm]{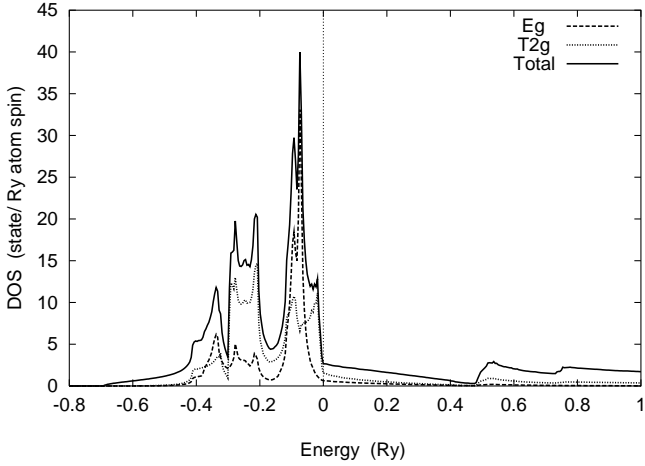}%
\caption{\label{ldados}
Densities of states (DOS) calculated by the LDA and TB-LMTO method.
Dashed curve: local DOS for e${}_{g}$ electrons, dotted curve: local
 DOS for t${}_{2g}$ electrons, solid curve: total DOS consisting of 4$s$,
 4$p$, and 3$d$ orbitals 
}
\end{figure}
We have solved the dynamical CPA equation for the bcc Fe 
using the expressions (\ref{cohg2}) and (\ref{gapprox}).  
The densities of states (DOS) for 3$d$, 4$s$, and 4$p$ states were 
calculated by using von Barth-Hedin LDA potential.  
The total DOS and the $d$ DOS for e${}_{g}$ and t${}_{2g}$ electrons 
are shown in Fig.~\ref{ldados}.
Single-particle excitation spectra have been calculated by using the 
Pad\'{e} numerical analytic contribution.

\begin{figure}
\includegraphics[width=9cm]{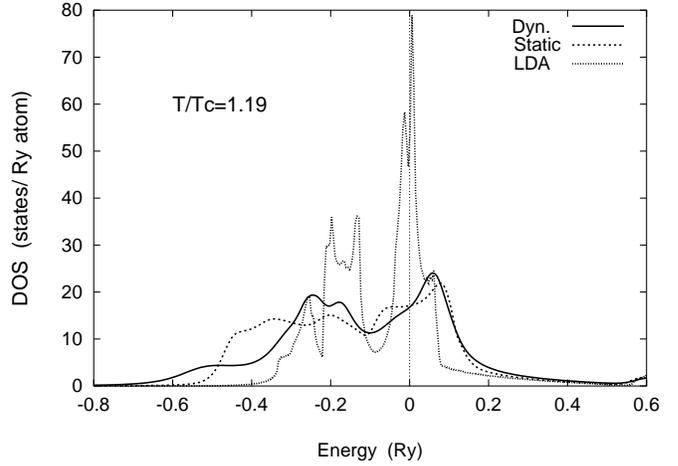}%
\caption{\label{fedosp}
Single particle excitation spectra (DOS) for $d$ electrons 
in the paramagnetic Fe.
Results for the LDA, the static approximation, and the 2nd-order
 dynamical CPA are shown by the dotted curve, the dashed
 curve, and the solid curve, respectively.  
}
\end{figure}

Figure \ref{fedosp} shows the calculated $d$ DOS of paramagnetic Fe
at $T/T_{\rm C}=1.19$ as the single-particle excitation spectra.
The DOS in the static approximation is broadened as compared with
the LDA result in the nonmagnetic state because of the strong thermal 
spin fluctuations.   
The dynamical charge and spin fluctuations produce a
satellite peak around $\omega=-0.5$ Ry (=$ - 6.8$ eV), and suppress 
the band broadening by about 22\% as compared with the static one.  
The existence of the satellite peak is consistent with the previous 
results of the ground-state calculations~\cite{unger94} 
as well as those at finite temperatures~\cite{lich01}.
The $d$ band width in the present calculations, though it is strongly
reduced as compared with the static one, is
comparable to that of the LDA calculations, while the XPS experiments
suggest the 10\% reduction of the width as compared with the LDA 
results, and no dip at $\omega=-0.1$ Ry~\cite{eastman80}.
These inconsistencies may be attributed to
an overestimate of the local exchange splitting above $T_{\rm C}$.

\begin{figure}
\includegraphics[width=9cm]{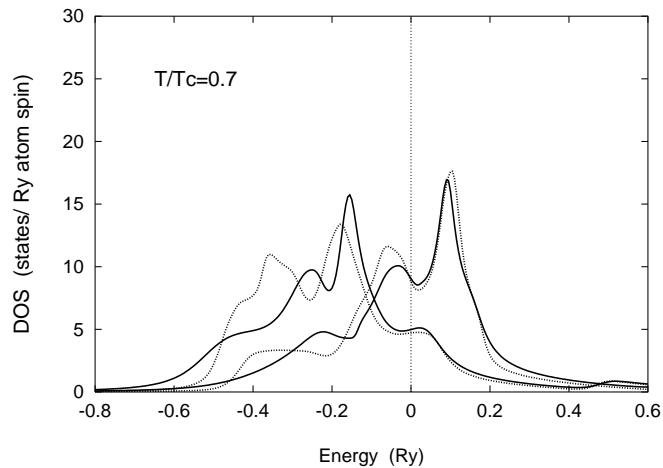}%
\caption{\label{fedosf.7}
Up and down $d$ DOS in the ferromagnetic Fe at $T/T_{\rm C}=0.7$.
Results for the static approximation are shown by the dotted curves.
}
\end{figure}

Below the Curie temperature, the up and down DOS are split as shown in
Fig.~\ref{fedosf.7}.  In the up-spin band, the satellite peak at 
$\omega = -0.45$ Ry remains, and the quasiparticle bands 
at $\omega \approx -0.2$ Ry shifts to the Fermi level as compared with
those in the static approximation, showing the band
narrowing.  The satellite peak for the down-spin band disappears because
of a large value of $|{\rm Im} \Sigma_{L\sigma}(z)|$ in this energy region.
These behaviors are consistent with recent QMC calculations without
transverse spin fluctuations~\cite{lich01}.

\begin{figure}
\includegraphics[width=9cm]{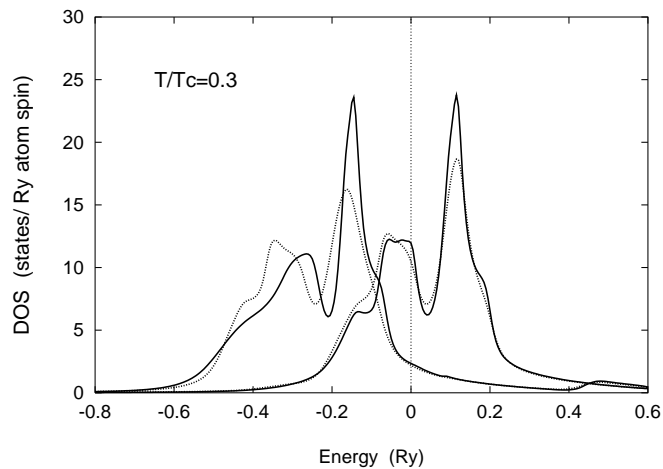}%
\caption{\label{fedosf.3}
Up and down $d$ DOS in the ferromagnetic Fe at $T/T_{\rm C}=0.3$.
}
\end{figure}

It is not easy to calculate the DOS at low temperatures in the QMC
calculations.  The present approach allows us to
investigate the DOS even at low temperatures.  Figure \ref{fedosf.3} 
shows the DOS at $T/T_{\rm C}=0.3$.  The DOS in the static
approximation approaches to the Hartree-Fock one with decreasing
temperature, but are still broadened at this temperature by 
thermal spin fluctuations.
Dynamical terms suppress the thermal spin fluctuations and develops the
quasiparticle states, so that sharp peaks of e${}_{\rm g}$ electrons
appear at $\omega = \pm 0.15$ Ry in the DOS.
The present calculations reduce to the 2nd-order
perturbation theory at $T=0$, so that the DOS in Fig. \ref{fedosf.3} is
close to those obtained at the zero temperature by Drchal 
{\it et. al.}~\cite{drchal99}.

\begin{figure}
\includegraphics[width=9cm]{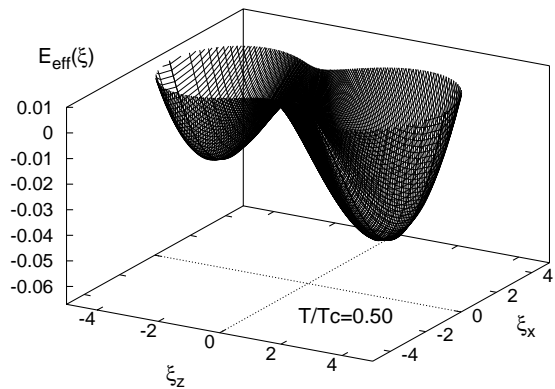}%
\caption{\label{feexi}
Effective potential in the ferromagnetic Fe at $T/T_{\rm C}=0.5$ on 
the $\xi_{x}$-$\xi_{z}$ plane. 
}
\end{figure}

\begin{figure}
\includegraphics[width=9cm]{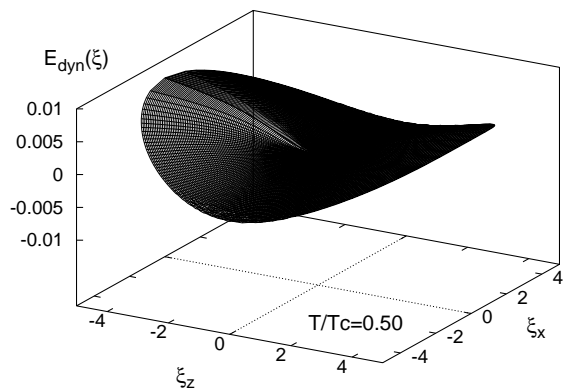}%
\caption{\label{feedyn}
Dynamical contribution to effective potential in the ferromagnetic Fe 
at $T/T_{\rm C}=0.5$ on the $\xi_{x}$-$\xi_{z}$ plane. 
}
\end{figure}

The effective potential determines the behavior of magnetic moments.
Figure \ref{feexi} shows the potential for Fe below $T_{\rm C}$.
It has double minima along $z$ axis,
and monotonically increases with increasing
$\xi_{\perp}=\sqrt{\xi^{2}_{x}+\xi^{2}_{y}}$.  
(Note that the effective potential is spherical on the $xy$ plane: 
$E_{\rm eff}(\xi_{z}, \xi_{\perp})$).
The dynamical contribution $E_{\rm dyn}(\boldsymbol{\xi})$ to the 
effective potential is given in Fig.~\ref{feedyn}.
The dynamical part shows a 'butterfly' structure; it increases along the
$z$ axis with increasing the amplitude $|\boldsymbol{\xi}|$, while it
decreases on the $xy$ plane.  This implies that the dynamical effects
reduce the longitudinal amplitude of spin fluctuations, and enhance the
transverse spin fluctuations.  In fact, we find 6\% reduction of
$\sqrt{\langle \xi^{2}_{z} \rangle}$ and 6\% enhancement of 
$\sqrt{\langle \xi^{2}_{\perp} \rangle}$ at $T/T_{\rm C}=1.19$.

\begin{figure}
\includegraphics[width=9cm]{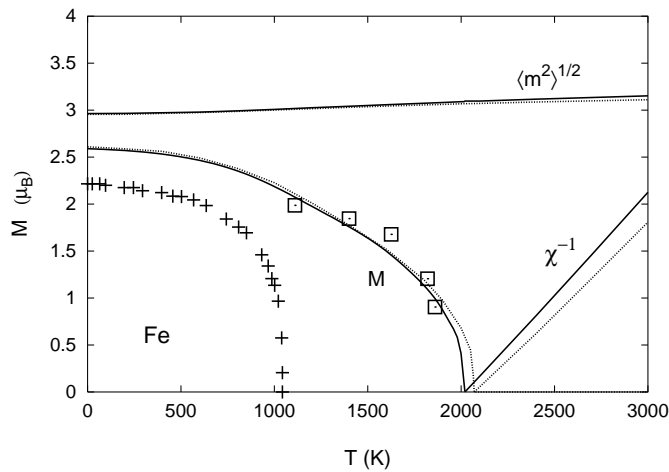}%
\caption{\label{femt}
Calculated magnetization ($M$), inverse susceptibility ($\chi^{-1}$), and 
amplitude of local moment ($\langle {\bf m}^{2} \rangle^{1/2}$) 
as a function of temperature ($T$) in Fe.
The dynamical results are shown by the solid curves.
Results in the static approximation are shown by dotted curves.
Magnetization calculated by the DMFT without transverse spin 
fluctuations~\cite{lich01} is also shown by open squares.
Experimental data of magnetization~\cite{potter34} are shown by $+$. 
Note that the absolute values of the DMFT magnetization are not given in
 Ref. 29.  Thus they are plotted here by assuming that the
 extrapolated value at $T=0$ agrees with the experimental one.
}
\end{figure}

Magnetic properties of Fe are summarized in Fig.~\ref{femt}.  Both
static and dynamical calculations yield the Curie-Weiss susceptibility.
Calculated effective Bohr magneton numbers are 3.1 $\mu_{\rm B}$ in the
static approximation and 3.0 $\mu_{\rm B}$ in the
dynamical calculations, respectively, being in good agreement with the
experimental value 3.2 $\mu_{\rm B}$~\cite{fallot44}.
Calculated Curie temperature is 2020 K (2070 K) in the 2nd-order
dynamical calculations (the static approximation).  They are much
smaller than the Hartree-Fock value 12200 K, but still twice as large as
the experimental value (1040 K)~\cite{arrot67}.
The present results are comparable to the QMC result of calculations 
without transverse spin fluctuations (1900 K)~\cite{lich01}.
The reduction of $T_{\rm C}$ due to dynamical corrections is 50 K, which
is rather small.
Dynamical effects in general reduce the
magnetic energy, but also reduce the magnetic entropy of the static
approximation. Both effects are competitive to each other, resulting 
in the reduction of $T_{\rm C}$ by 50 K in the case of Fe.

The magnetization increases with decreasing temperature, and reach the
Hartree-Fock value 2.61 $\mu_{\rm B}$ at $T=0$ K in the static
approximation.  The latter is overestimated as compared with the
experimental value 2.216 $\mu_{\rm B}$~\cite{danan68}.  
The 2nd-order dynamical CPA
calculations yield $M=2.59$ $\mu_{\rm B}$ (extrapolated value); 
the calculations
hardly reduce the ground-state magnetization as seen in
Fig.~\ref{femt}. 
One has to take into account the higher-order electron-electron 
scattering effects as found in the low-density approximation~\cite{kana63}
to reduce the magnetization.  
The amplitude of local magnetic moment was calculated by means of eq. 
(\ref{avm2l2}).  The results are plotted in the same figure.  
Because of the strong Coulomb interaction, it hardly changes with 
increasing temperature.  The dynamical fluctuations enhance the
amplitude $\sqrt{\langle \hat{\boldsymbol{m}}^{2} \rangle}$ by 1\%, 
and reduce the $d$ charge fluctuations 
$\sqrt{\langle (\delta \hat{n}_{d})^{2} \rangle}$ by 5\% at 
$T/T_{\rm C}=1.2$.
\begin{figure}[h]
\includegraphics[width=9cm]{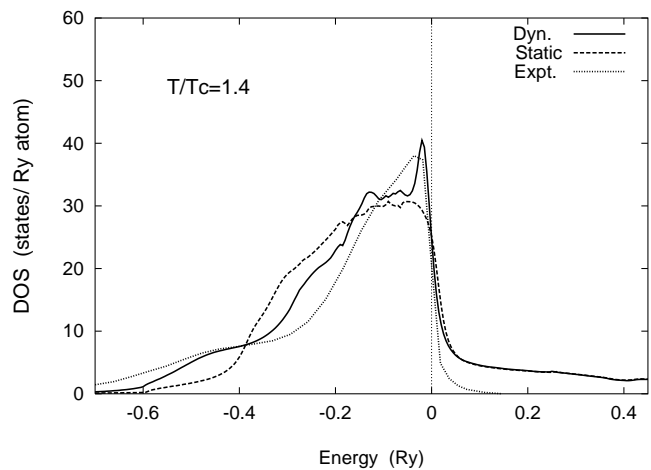}%
\caption{\label{nidosp}
Calculated DOS in the paramagnetic Ni.
Solid curve: 2nd-order dynamical CPA, dashed curve:
 static approximation, dotted curve: XPS data~\cite{himpsel79}.  
}
\end{figure}
\begin{figure}[h]
\includegraphics[width=9cm]{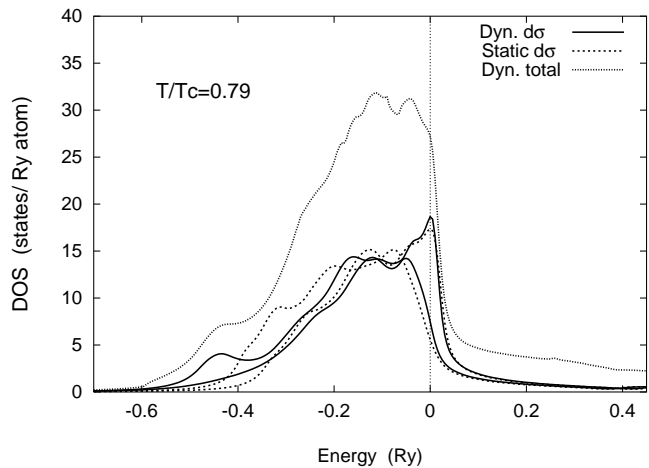}%
\caption{\label{nidosf}
Calculated DOS in the ferromagnetic Ni.
Solid curves : spin-polarized $d$ DOS in the 2nd-order dynamical CPA, 
dashed curves : spin-polarized $d$ DOS in the static approximation,
 and dotted curve: total DOS in the 2nd-order dynamical CPA.  
}
\end{figure}

We have also calculated the electronic and magnetic properties
of the fcc Ni at finite temperatures.  
Figure \ref{nidosp} shows the DOS in the paramagnetic state.  In the
static approximation, the details of the structure are smeared by
thermal spin fluctuations and the 
$d$ band width is broadened by about 0.1 Ry.
The dynamical effects suppress the thermal spin fluctuations and develop
the quasiparticle states.  Reduction of the quasiparticle band width
is 17\%  as compared with that of the static approximation.  
Furthermore we 
find the satellite peak at $\omega=-0.45$ Ry.  These results explain
well the XPS data~\cite{himpsel79} as shown in Fig.~\ref{nidosp}.

Below $T_{\rm C}$, the peak of the down-spin band is on the Fermi level,
as shown in Fig. \ref{nidosf}.
On the other hand, the top of the up-spin $d$ band is away from the
Fermi level, so that the peak is weakened due to larger damping of the
quasiparticle states.
The satellite peak for the down-spin band disappears due to strong
incoherent scatterings around $\omega=-0.35$ Ry, while the satellite peak
for the up-spin band is enhanced at $\omega=-0.45$ Ry. 

\begin{figure}
\includegraphics[width=9cm]{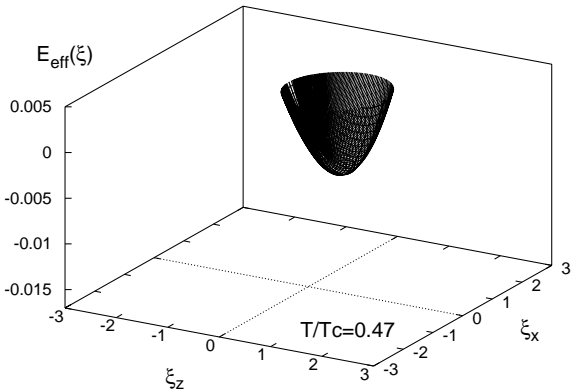}%
\caption{\label{niexi}
Effective potential in the ferromagnetic Ni at $T/T_{\rm C}=0.47$
on the $\xi_{x}$-$\xi_{z}$ plane.
}
\end{figure}

The effective potential for Ni shows a single minimum structure as shown
in Fig.~\ref{niexi}.  The minimum position shifts to the origin with
increasing temperature.  This should be contrasted to the case of Fe, 
in which the effective potential has a double minimum structure even 
above $T_{\rm C}$ as shown in Fig.~\ref{feexi}, 
and the paramagnetic state is realized by changing 
the energy difference between the two minima.
The dynamical potential $E_{\rm dyn}(\boldsymbol{\xi})$ in Ni has a 
butterfly structure as in the case of Fe, but it is highly 
asymmetric along the $z$ axis in the ferromagnetic state 
so that considerable reduction of the magnetization due to
dynamical corrections occurs.  We
find the reduction of $\sqrt{\langle \xi^{2}_{z} \rangle}$ by 5.0\%, and
the enhancement of $\sqrt{\langle \xi^{2}_{\perp} \rangle}$ by 1.5\% at
$T/T_{\rm C}=1.3$.
\begin{figure}
\includegraphics[width=9cm]{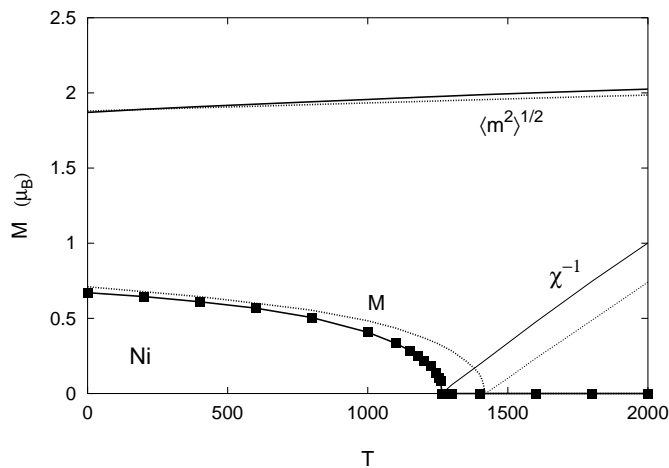}%
\caption{\label{nimt}
Magnetization, inverse susceptibility, and amplitude of local moment as
 a function of temperature in Ni.
The dynamical results are shown by the solid curves, while the results 
in the static calculation are shown by dotted curves. 
}
\end{figure}

The magnetic moment and the inverse susceptibility calculated from 
the effective
potential are presented in Fig.~\ref{nimt} as a function of
temperature.  The susceptibility follows the Curie-Weiss law.  Both the
static and dynamical calculations yield the effective Bohr magneton
number 1.2 $\mu_{\rm B}$, which should be compared with the experimental
value 1.6 $\mu_{\rm B}$~\cite{sucksmith38}.  
Calculated Curie temperature in Ni is 1260 K
(1420 K) in the 2nd-order dynamical (static) calculations.
These values are much smaller than the Hartree-Fock value 4950 K, but
are still twice as large as the experimental value 630 K~\cite{arrot67}.

The magnetization increases with decreasing temperature below 
$T_{\rm C}$.  Extrapolated value at $T=0$ is 0.67 (0.71) $\mu_{\rm B}$ 
in the 2nd-order dynamical (static) calculations.  These values are
considerably larger than the experimental one 
(0.62 $\mu_{\rm B}$)~\cite{danan68}. 
The amplitude of Ni local moment 
slightly increases with increasing temperature
and hardly shows anomaly at $T_{\rm C}$.  The 2nd order dynamical
corrections to the amplitude of local moment and the local charge 
fluctuations are
small; the enhancement of $\sqrt{\langle \hat{\boldsymbol{m}}^{2} \rangle}$
is only 1.8 \% and the reduction of the $d$ charge fluctuations
$\sqrt{\langle (\delta \hat{n}_{d})^{2} \rangle}$ is 5.9 \% at 
$T/T_{\rm C}=1.3$.

\section{Summary}
We have developed the dynamical CPA on the basis of the LDA+TB-LMTO
Hamiltonian towards realistic calculations of the
itinerant electron system.
The theory is a direct extension of the single-site theory
developed by Cyrot, Hubbard, Hasegawa, and Kakehashi, to the degenerate
band case.
It is based on the functional integral method which transforms the
interacting electron system into an independent electron system with
time-dependent random charge and exchange potentials.  Using the method, 
we have taken into account the spin fluctuations as well as charge
fluctuations in the degenerate band
system.  We have then introduced an effective medium
$\Sigma_{L\sigma}(i\omega_{n})$, and derived the self-consistent dynamical
CPA equation for the medium, making use of a single-site
approximation. 

We adopted the harmonic approximation (HA) to treat the functional 
integrals in the dynamical CPA.
The HA describes the dynamical effects from the weak- to the strong- 
Coulomb interaction regime.
The approximation allows us to obtain analytical expressions of the
physical quantities, and takes into account the dynamical corrections
successively starting from a high-temperature approximation ({\it i.e.},
the static approximation).  
We can calculate the excitation spectra as well as the thermodynamic 
quantities even at low temperatures
using the HA because we obtained their analytic expressions.

We have investigated the dynamical effects in Fe and Ni within the
2nd-order dynamical CPA, and have shown that the 2nd-order dynamical 
corrections much improve the single-particle excitation spectra in 
these systems.
The static approximation broadens the DOS due to thermal spin
fluctuations at finite temperatures.  The dynamical effects suppress the
thermal spin fluctuations and create the quasiparticle states with
narrow band width near the Fermi level.  Furthermore, the correlations
create the satellite peak at 6 eV below the Fermi level in both Fe and
Ni.  The XPS data in the paramagnetic Ni is well explained by the
present theory.

We verified that the dynamical CPA
yields the Curie-Weiss susceptibilities.  Calculated effective Bohr 
magneton numbers, 3.0 $\mu_{\rm B}$ for Fe and 1.2 $\mu_{\rm
B}$ for Ni, explain the experimental data quantitatively or
semiquantitatively.  Calculated Curie
temperatures, 2020 K for Fe and 1260 K for Ni, are however 
overestimated by a factor of two. 
Extrapolated values of the ground state magnetization, 2.59 $\mu_{\rm
B}$ for Fe and 0.67 $\mu_{\rm B}$ for Ni, are also 
overestimated considerably as compared with the experimental ones 
(2.22 $\mu_{\rm B}$ for Fe and 0.62 $\mu_{\rm B}$ for Ni).  

We found that the static approximation provides us with a good 
starting point to
calculate finite-temperature magnetic properties of Fe and Ni, but 
the dynamical calculations to go beyond the static approximation 
have been limited to the second-order dynamical CPA in the present 
work.  Overestimate of the ground-state 
magnetization and the Curie temperature should be reduced by
taking into account the higher-order dynamical fluctuations.
Further improvements of the dynamical CPA theory are left for future 
investigations.
\vspace{10mm}

\section*{Acknowledgment}
We would like to express our sincere thanks to Dr. Ove Jepsen for
sending us the Stuttgart TB-LMTO program and fruitful advice on how to
install the program on our computer.
\vspace{10mm}

\appendix
\section{Expression of $\hat{D}^{(n)}_{\{ \alpha\gamma \}}(\nu,k,m)$}   
\vspace{10mm}

We calculate in this appendix the coefficients 
$\hat{D}^{(n)}_{\{ \alpha\gamma \}}(\nu,k,m)$
in the $n$-th order expansion of the determinant $D_{\nu}(k,m)$ with
respect to the dynamical potential $v_{\alpha}(\nu,m)$.

Let us rewrite $D_{\nu}(k,m)$ defined by eq. (\ref{dnukm}) as follows by
making use of the Laplace expansion.

\begin{eqnarray}
D = |a^{(0)}|\overline{D}_{20}D_{20} 
+ \sum_{\alpha\gamma} (a^{(0)})_{\alpha\gamma} \Delta_{\gamma\alpha} 
+ \overline{D}_{10}D_{10} \ ,
\label{ddet}
\end{eqnarray}
\begin{eqnarray}
\Delta_{11} = - (\overline{D}_{20}- \overline{D}_{24})(D_{20}- D_{24})
- \overline{D}_{23}D_{23} \ ,
\label{d11}
\end{eqnarray}
\begin{eqnarray}
\Delta_{12} = - \overline{D}_{23}(D_{20}- D_{21})
- (\overline{D}_{20}- \overline{D}_{24})D_{22} \ ,
\label{d12}
\end{eqnarray}
\begin{eqnarray}
\Delta_{21} = - \overline{D}_{22}(D_{20}- D_{24})
- (\overline{D}_{20}- \overline{D}_{21})D_{23} \ ,
\label{d21}
\end{eqnarray}
\begin{eqnarray}
\Delta_{22} = - (\overline{D}_{20}- \overline{D}_{21})(D_{20}- D_{21})
- \overline{D}_{22}D_{22} \ .
\label{d22}
\end{eqnarray}

In the above equations, we have omitted the suffixes $\nu$, $k$, $m$ 
for simplicity, and $|a^{(0)}|$ denotes the determinant of the 
$2 \times 2$ matrix $a_{k}(\nu,m)$.
$\{ D_{n\alpha} \}$ at the r.h.s. of eqs. (\ref{ddet}-\ref{d22}) are
defined by 
\begin{eqnarray}
D_{n\alpha} = \left| 
\begin{array}{@{\,}cccccc@{\,}}
\ b^{(n-1)}_{\alpha} & 1           &     &            & 0  &   \\
a^{(n)}_{\alpha}     & 1           & \ \ 1   &            &    &   \\
                & a^{(n+1)}   & \ \ 1   & \ \ 1      &    &   \\
                &             &     & \ \ \ddots &    &   \\
\ \ 0           &             &     &            &    &   \\
\end{array}
\right| \ ,
\label{dn1}
\end{eqnarray}
\begin{eqnarray}
\overline{D}_{n\alpha} = \left| 
\begin{array}{@{\,}cccccc@{\,}}
\ \overline{b}^{(n-1)}_{\alpha} & \overline{a}^{(n)}_{\alpha} & & & 0  &   \\
1     & 1           & \ \ \overline{a}^{(n+1)}   &            &    &   \\
                & 1  & 1 \ \  & \hspace*{-3mm} \overline{a}^{(n+2)} & &   \\
                &             &     & \hspace*{-5mm} \ddots &    &   \\
\ \ 0           &             &     &            &    &   \\
\end{array}
\right| \ .
\label{dn2}
\end{eqnarray}
Here $a^{(n)}$ ($\overline{a}^{(n)}$) stands for $a_{n\nu+k}(\nu,m)$ 
($a_{-n\nu+k}(\nu,m)$).  $a^{(n)}_{\alpha}$,
$b^{(n)}_{\alpha}$, $\overline{a}^{(n)}_{\alpha}$, and 
$\overline{b}^{(n)}_{\alpha}$ are defined by   
$a^{(n)}_{0}=a^{(n)}$, $b^{(n)}_{0}=1$, 
$\overline{a}^{(n)}_{0}=a^{(-n)}$, $\overline{b}^{(n)}_{0}=1$, and 
for $\alpha=1 \sim 4$,
\begin{eqnarray}
a^{(n)}_{1}\!=\!a^{(n)}_{2}\!=\! \left(
\begin{array}{@{\,}cc@{\,}}
0 & a^{(n)}_{12}  \\
0 & a^{(n)}_{22}  \\
\end{array}
\right) , 
a^{(n)}_{3}\!=\!a^{(n)}_{4}\!= \!\left(
\begin{array}{@{\,}cc@{\,}}
0 \!\!\!\!& a^{(n)}_{11}  \\
0 \!\!\!\!& a^{(n)}_{21}  \\
\end{array}
\right) ,
\label{defan}
\end{eqnarray}
\begin{eqnarray}
b^{(n)}_{1} &=& \left(
\begin{array}{@{\,}cc@{\,}}
a^{(n)}_{11} & 0  \\
a^{(n)}_{21} & 1  \\
\end{array}
\right)  , \ \ 
b^{(n)}_{2} = \left(
\begin{array}{@{\,}cc@{\,}}
a^{(n)}_{12} & 0  \\
a^{(n)}_{22} & 1  \\
\end{array}
\right)  , \nonumber \\
& & \hspace{-15mm}
b^{(n)}_{3} = \left(
\begin{array}{@{\,}cc@{\,}}
a^{(n)}_{11} & 1  \\
a^{(n)}_{21} & 0  \\
\end{array}
\right)  , \ \ 
b^{(n)}_{4} = \left(
\begin{array}{@{\,}cc@{\,}}
a^{(n)}_{12} & 1  \\
a^{(n)}_{22} & 0  \\
\end{array}
\right) \ ,
\label{defbn}
\end{eqnarray}
\begin{eqnarray}
\overline{a}^{(n)}_{1}&=&\overline{a}^{(n)}_{2}= \left(
\begin{array}{@{\,}cc@{\,}}
0 & 0  \\
a^{(-n)}_{21} & a^{(-n)}_{22}  \\
\end{array}
\right) \ , \nonumber \\
& &  \hspace{-17mm}
\overline{a}^{(n)}_{3}\ \ =\ \ \overline{a}^{(n)}_{4}= \left(
\begin{array}{@{\,}cc@{\,}}
0 & 0  \\
a^{(-n)}_{11} & a^{(-n)}_{12}  \\
\end{array}
\right) \ ,
\label{defbaran}
\end{eqnarray}
\begin{eqnarray}
\overline{b}^{(n)}_{1} \!\! &=& \!\! \left(
\begin{array}{@{\,}cc@{\,}}
a^{(-n)}_{11} & \hspace{-5mm} a^{(-n)}_{12}  \\
\hspace{-4mm} 0 & \hspace{-9mm} 1  \\
\end{array}
\!\! \right) , \, 
\overline{b}^{(n)}_{2} \!\! = \!\! \left(
\begin{array}{@{\,}cc@{\,}}
a^{(-n)}_{21} & \hspace{-5mm} a^{(-n)}_{22}  \\
\hspace{-4mm} 0 & \hspace{-9mm} 1  \\
\end{array}
\right) , \, \nonumber \\
& &  \hspace{-15mm}
\overline{b}^{(n)}_{3}  =  \left(
\begin{array}{@{\,}cc@{\,}}
a^{(-n)}_{11} & \hspace{-5mm} a^{(-n)}_{12}  \\
\hspace{-4mm} 1 & \hspace{-9mm} 0  \\
\end{array}
\right) , \,
\overline{b}^{(n)}_{4} \!\! = \!\! \left(
\begin{array}{@{\,}cc@{\,}}
a^{(-n)}_{21} & \hspace{-5mm} a^{(-n)}_{22}  \\
\hspace{-4mm} 1 & \hspace{-9mm} 0  \\
\end{array}
\right) . 
\label{defbarbn}
\end{eqnarray}

It should be noted that eq. (\ref{ddet}) is calculated from a set 
($D_{10}, D_{20}, D_{21}, D_{22}, D_{23}, D_{24}$).
Thus we define $\boldsymbol{D}^{(n)}$ by 
${}^{\rm t}\boldsymbol{D}^{(n)} = (D_{n0}, D_{n+1\, 0}, D_{n+1\, 1},
D_{n+1\, 2}, D_{n+1\, 3}, D_{n+1\, 4})$.
By making use of the Laplace expansion, we can derive a recursion
relation as follows.
\begin{eqnarray}
\boldsymbol{D}^{(n)} = (c_{0} + c^{(n)}_{1} + c^{(n)}_{2}) 
\boldsymbol{D}^{(n+2)} \ .
\label{drecur}
\end{eqnarray}
Here $(c_{0})_{ij}=\delta_{i1}\delta_{j1}+\delta_{i2}\delta_{j1}$, and 
\begin{eqnarray}
\hspace*{-5mm}c^{(n)}_{1} \!= \!\left( 
\begin{array}{@{\,}cccccc@{\,}}
\!\!\!-a^{(n)}_{11}\!\!\!-\!a^{(n)}_{22}\!\!\!\!\!\!\!\! & \!-a^{(n\!+\!1)}_{11} \!\!\!\!\!\!\!\! -a^{(n\!+\!1)}_{22} \!\!\!\!\!\!& 
a^{(n\!+\!1)}_{22} \!\!\!\!\!\!\!\!& -a^{(n\!+\!1)}_{21}\!\!\!\!\!\! & -a^{(n\!+\!1)}_{12}\!\!\!\!\!\! & a^{(n\!+\!1)}_{11}\!\!\!\!  \\
0      & \!-a^{(n\!+\!1)}_{11} \!\!\!\!\!\!\!\! -a^{(n\!+\!1)}_{22}\!\!\!\!\!\! & a^{(n\!+\!1)}_{22} \!\!\!\!\!\!\!\!& 
-a^{(n\!+\!1)}_{21}\!\!\!\!\!\! & -a^{(n\!+\!1)}_{12}\!\!\!\!\!\! & a^{(n\!+\!1)}_{11}\!\!\!\!  \\
a^{(n)}_{11} & 0 & 0 & 0 & 0 & 0  \\
a^{(n)}_{12} & 0 & 0 & 0 & 0 & 0  \\
a^{(n)}_{21} & 0 & 0 & 0 & 0 & 0  \\
a^{(n)}_{22} & 0 & 0 & 0 & 0 & 0  \\
\end{array}
\right)\!\!, \hspace{1mm}
\label{cn1}
\end{eqnarray}
\begin{eqnarray}
\hspace*{-12mm}
c^{(n)}_{2} \!\!\!=\!\!\! \left( 
\begin{array}{@{\,}cccccc@{\,}}
\hspace*{-2mm}|a^{(n)}| \hspace{-1mm}& 
\hspace{-4mm} c^{(n,n+1)}_{12} & 
\hspace{-5mm} -c^{(n,n+1)}_{11221221} & 
\hspace{-5mm} c^{(n,n+1)}_{22212122} & 
\hspace{-5mm} c^{(n,n+1)}_{11121211} & 
\hspace{-5mm} -c^{(n,n+1)}_{22112112}\hspace{-2mm} \\
\hspace*{-2mm}0 \hspace{-2mm} & \hspace{-4mm} |a^{(n\!+\!1)}| & \hspace{-4mm} 0 & 
\hspace{-4mm} 0 & \hspace{-4mm} 0 & \hspace{-5mm} 0\hspace{-2mm}  \\
\hspace*{-3mm}0 \hspace{-2mm} & 
\hspace{-4mm} -c^{(n,n+1)}_{11222112} & 
\hspace{-5mm} a^{(n)}_{11}a^{(n\!+\!1)}_{22} & 
\hspace{-5mm} a^{(n)}_{21}a^{(n\!+\!1)}_{22} & 
\hspace{-5mm} -a^{(n)}_{11}a^{(n\!+\!1)}_{12} & 
\hspace{-5mm} -a^{(n)}_{21}a^{(n\!+\!1)}_{12}\hspace{-2mm}  \\
\hspace*{-2mm}0 \hspace{-2mm} & 
\hspace{-4mm} c^{(n,n+1)}_{22121222} & 
\hspace{-5mm} a^{(n)}_{12}a^{(n\!+\!1)}_{22} & 
\hspace{-5mm} a^{(n)}_{22}a^{(n\!+\!1)}_{22} & 
\hspace{-5mm} -a^{(n)}_{12}a^{(n\!+\!1)}_{12} & 
\hspace{-5mm} -a^{(n)}_{22}a^{(n\!+\!1)}_{12}\hspace{-2mm}  \\
\hspace*{-2mm}0 \hspace{-2mm} &  
\hspace{-5mm} c^{(n,n+1)}_{11212111} & 
\hspace{-5mm} -a^{(n)}_{11}a^{(n\!+\!1)}_{21} & 
\hspace{-5mm} -a^{(n)}_{21}a^{(n\!+\!1)}_{21} & 
\hspace{-5mm} a^{(n)}_{11}a^{(n\!+\!1)}_{11} & 
\hspace{-5mm} a^{(n)}_{21}a^{(n\!+\!1)}_{11}\hspace{-2mm}  \\
\hspace*{-2mm}0 \hspace{-2mm} &  
\hspace{-5mm} -c^{(n,n+1)}_{22111221} & 
\hspace{-5mm} -a^{(n)}_{12}a^{(n\!+\!1)}_{21} & 
\hspace{-5mm} -a^{(n)}_{22}a^{(n\!+\!1)}_{21} & 
\hspace{-5mm} a^{(n)}_{12}a^{(n\!+\!1)}_{11} & 
\hspace{-5mm} a^{(n)}_{22}a^{(n\!+\!1)}_{11}\hspace{-2mm}  \\
\end{array}
\right) \!\!. \
\label{cn2}
\end{eqnarray}
Here 
$c^{(n,n+1)}_{\alpha\beta\gamma\delta
\alpha^{\prime}\beta^{\prime}\gamma^{\prime}\delta^{\prime}} = 
a^{(n)}_{\alpha\beta}
a^{(n+1)}_{\gamma\delta} - 
a^{(n)}_{\alpha^{\prime}\beta^{\prime}}
a^{(n+1)}_{\gamma^{\prime}\delta^{\prime}}$, and
$c^{(n,n+1)}_{12}=c^{(n,n+1)}_{11221221}+c^{(n,n+1)}_{22112112}+|a^{(n+1)}|$.
Note that $c_{0}$, $c^{(n)}_{1}$, and $c^{(n)}_{2}$ matrices are of the
0-th order, the first order, and the second order with respect to the
dynamical potential $v$, respectively.

Using the relation $c_{0}^{2}=c_{0}$ and $D_{\infty 0}=1$, we obtain
the relation.
\begin{eqnarray}
\boldsymbol{D}^{(m)} = \boldsymbol{E}_{2}
+ \sum_{n=0}^{\infty} c_{0}^{n} 
(\boldsymbol{D}^{(m+2n)}-c_{0}\boldsymbol{D}^{(m+2n+2)}) \ .
\label{drecur2}
\end{eqnarray}
Here ${}^{t}\boldsymbol{E}_{2} = (1,1,0,0,0,0)$.

Substituting eq. (\ref{drecur}) into eq. (\ref{drecur2}), and using the
recursion relation successively, we reach the expansion of 
$\boldsymbol{D}^{(1)}$ with respect to the dynamical potential $v$.
\begin{eqnarray}
\boldsymbol{D}^{(1)}\!\! &=&\!\! \boldsymbol{E}_{2} \hspace{20mm}\nonumber \\
& & \hspace{-20mm}
+ \!\!\sum_{n=1}^{\infty} \!\sum_{k \ge n/2}^{n} \!\sum_{l_{k}=0}^{\infty}
\!\sum_{l_{k\!-\!1}=0}^{l_{k}} \!\!\!\!\!\cdots \!\!\!\sum_{l_{1}=0}^{l_{2}}
\hspace{-6mm}
\sum_{\ \ \ \ \ i_{1}\!+\! \cdots \!+\!i_{k}\!=\!n}  \hspace{-7mm} 
c_{0}^{l_{1}}c_{i_{1}}^{2l_{\!1}\!+\!1} \!\!\!\!\cdots \!\!
c_{0}^{l_{k}\!-\!l_{k-1}}c_{i_{k}}^{2l_{k}\!+\!2k\!-\!1}
\boldsymbol{E}_{2} , \hspace{10mm}
\label{d1exp}
\end{eqnarray}
where $i_{1} \cdots i_{k}$ take a value 1 or 2.

In the same way, we obtain the expansion of 
$\overline{\boldsymbol{D}}^{(1)}$ as
\begin{eqnarray}
\overline{\boldsymbol{D}}^{(1)}\!\!\! &=& \!\!\!\boldsymbol{E}_{2} \nonumber \\
& & \hspace{-20mm}
+\!\! \sum_{n=1}^{\infty}\! \sum_{k \ge n\!/\!2}^{n} \!\sum_{l_{k}\!=\!0}^{\infty}
\sum_{l_{k\!-\!1}=0}^{l_{k}} \!\!\!\!\cdots\!\!\!\! \sum_{l_{1}=0}^{l_{2}}
\hspace{-6mm}\sum_{\ \ \ \ \ i_{1} \!+\! \cdots \!+\!i_{k}\!=\!n} \hspace{-7mm} 
c_{0}^{l_{1}}\overline{c}_{i_{1}}^{2l_{1}\!+\!1} \!\!\!\cdots 
c_{0}^{l_{k}\!-\!l_{k\!-\!1}}\overline{c}_{i_{k}}^{2l_{k}\!+\!2k\!-\!1}
\!\!\boldsymbol{E}_{2} .  \hspace{10mm}
\label{d1barexp}
\end{eqnarray}
Here $\overline{c}^{(n)}_{1}$ and $\overline{c}^{(n)}_{2}$ are defined
by $c_{1}$ and $c_{2}$ in which $\{ a^{(n)} \}$ have been replaced by
$\{ \overline{a}^{(n)} \}$. 

Substituting eqs. (\ref{d1exp}) and (\ref{d1barexp}) into
eq. (\ref{ddet}), we obtain the expansion of $D$ with respect to
dynamical potentials.
\begin{eqnarray}
D^{(n)}_{\nu}(k,m) \!\!\!&=& \!\!\!\sum_{n=0}^{\infty}
\sum_{\alpha_{1}\!\gamma_{1} \cdots \alpha_{n}\!\gamma_{n}}
\!\!\!\!\!\!\!\! v_{\alpha_{1}}(\nu,m)v_{\gamma_{1}}(-\nu,m) \cdots
\nonumber \\
& & \hspace{-10mm}\times 
v_{\alpha_{n}}(\nu,m)v_{\gamma_{n}}(-\nu,m) 
\hat{D}^{(n)}_{\{ \alpha\gamma \}}(\nu,k,m) .\hspace{10mm}
\label{dnnukm2}
\end{eqnarray}
Note that $\alpha_{n}$ and $\gamma_{n}$ take $0$, $x$, $y$, and $z$.

The first few terms of $\hat{D}^{(n)}_{\{ \alpha\gamma \}}(\nu,k,m)$ 
are expressed as follows. 
\begin{eqnarray}
\hat{D}^{(0)}(\nu,k,m) = 1 \ ,
\label{dhat0}
\end{eqnarray}
\begin{eqnarray}
\hat{D}^{(1)}_{\alpha\gamma}(\nu,k,m) = - \sum_{n=-\infty}^{\infty} 
\sum_{\sigma} \hat{a}_{\alpha\gamma}(\nu, m, n\nu+k)_{\sigma\sigma} ,
\label{dhat1}
\end{eqnarray}
\begin{eqnarray}
\hat{D}^{(2)}_{\alpha\gamma\alpha^{\prime}\gamma^{\prime}}(\nu,k,m) 
& = & \frac{1}{2} \, 
\hat{D}^{(1)}_{\alpha\gamma}(\nu,k,m) 
\hat{D}^{(1)}_{\alpha^{\prime}\gamma^{\prime}}(\nu,k,m)   \nonumber \\
& &   \hspace*{-35mm}
- \frac{1}{2} \!\!\!\sum_{n=-\infty}^{\infty}\!\!\! \left(\!\!
\sum_{\sigma} \hat{a}_{\alpha\!\gamma}(\nu, \!m, n\nu\!+\!k)_{\sigma\sigma}
\!\!\right)\!\!\!
\left( \!\!\sum_{\sigma} \!
\hat{a}_{\alpha^{\prime}\!\gamma^{\prime}}(\nu,\!m, n\nu\!+\!k)_{\sigma\sigma}
\!\!\!\right)      \nonumber \\
& &   \hspace*{-30mm}
+ \sum_{n=-\infty}^{\infty} \Big[
\hat{a}_{\alpha\gamma}(\nu, m, n\nu+k)_{\uparrow\uparrow}
\hat{a}_{\alpha^{\prime}\gamma^{\prime}}(\nu, m, n\nu+k)_{\downarrow\downarrow}
\nonumber \\
& & \hspace*{-22mm}
-\hat{a}_{\alpha\gamma}(\nu, m, n\nu+k)_{\downarrow\uparrow}
\hat{a}_{\alpha^{\prime}\gamma^{\prime}}(\nu, m, n\nu+k)_{\uparrow\downarrow} 
\nonumber \\
& &   \hspace*{-33mm}
- \sum_{\sigma\sigma^{\prime}} 
\hat{a}_{\alpha\!\gamma}(\nu, m, n\nu+k)_{\sigma\sigma^{\prime}}
\hat{a}_{\alpha^{\prime}\!\gamma^{\prime}}
(\nu, m, n\nu+k)_{\sigma^{\prime}\sigma}
\Big] .  \hspace{10mm}
\label{dhat2}
\end{eqnarray}
Here $\hat{a}_{\alpha\gamma}(\nu,m,n)$ is defined by
\begin{eqnarray}
\hat{a}_{\alpha\gamma}(\nu,m,n) &=&  \nonumber \\
& & \hspace{-25mm}
\left[
(1 + O_{1}\sigma_{x} + O_{2}\sigma_{y} + O_{3}\sigma_{z})
\,\bar{h}(m, n-\nu, n) \right]_{\alpha\gamma} . \hspace{10mm}
\label{ahat}
\end{eqnarray}
$O_{1}$, $O_{2}$, $O_{3}$, and $\bar{h}$ in eq. (\ref{ahat}) are 
$4 \times 4$ matrices defined by
\begin{eqnarray}
\hspace*{-3mm}
O_{1} \!\!\!\!\!&=&\!\!\!\!\! \left(
\begin{array}{@{\,}cc@{\,}}
\sigma_{x} & \!\!\!\! 0 \\
0 & \!\!\!\! -\sigma_{y}
\end{array}
\right) , \ 
O_{2} = \!\left(
\begin{array}{@{\,}cc@{\,}}
0 & \!\!\!\! \lambda^{\ast} + \lambda \sigma_{z} \\
\lambda + \lambda^{\ast} \sigma_{z} & \!\!\!\! 0
\end{array}
\right) ,  \nonumber \\
& &  \hspace{-12mm}
O_{3} = \left(
\begin{array}{@{\,}cc@{\,}}
0 & \!\!\!\! \lambda (\sigma_{x}+\sigma_{y}) \\
\lambda^{\ast} (\sigma_{x}+\sigma_{y}) & \!\!\!\! 0
\end{array}
\right) ,  \hspace{8mm}
\label{oi}
\end{eqnarray}
\begin{eqnarray}
\hspace{-10mm}
\bar{h}(m, n\!-\!\nu, n) \!\!\!\!& = &    \nonumber \\
& &    \hspace{-30mm}
\left(
\begin{array}{@{\,}cccc@{\,}}
\hspace{-1mm}e_{0}\!+\!e_{x}\!\!+\!e_{y}\!\!+\!e_{z}\hspace{-6mm}  & a^{(+)}_{x}\!\!-\!ib^{(-)}_{x}\hspace{-6mm} & 
a^{(+)}_{y}\!\!-\!ib^{(-)}_{y}\hspace{-6mm} & a^{(+)}_{z}\!\!-\!ib^{(-)}_{z}  \\
a^{(+)}_{x}\!\!+\!ib^{(-)}_{x}\hspace{-6mm} & e_{0}\!+\!e_{x}\!\!-\!e_{y}\!\!-\!e_{z}\hspace{-6mm} &  
b^{(+)}_{z}\!\!-\!ia^{(-)}_{z}\hspace{-6mm} & b^{(+)}_{y}\!\!+\!ia^{(-)}_{y} \\
a^{(+)}_{y}\!\!+\!ib^{(-)}_{y}\hspace{-6mm} & b^{(+)}_{z}\!\!+\!ia^{(-)}_{z}\hspace{-6mm} & 
e_{0}\!\!-\!e_{x}\!\!+\!e_{y}\!\!-\!e_{z}\hspace{-6mm}  & b^{(+)}_{x}\!\!-\!ia^{(-)}_{x} \\
a^{(+)}_{z}\!\!+\!ib^{(-)}_{z}\hspace{-6mm} & b^{(+)}_{y}\!\!-\!ia^{(-)}_{y}\hspace{-6mm} & 
b^{(+)}_{x}\!\!+\!ia^{(-)}_{x}\hspace{-6mm} & e_{0}\!\!-\!e_{x}\!\!-\!e_{y}\!\!+\!\!e_{z} \hspace{-1mm}    
\end{array}
\right) . \hspace{8mm}
\label{hbar}
\end{eqnarray}
Here
$\lambda = (1+i)/2$, and 
\begin{eqnarray}
e_{\alpha} = g^{\alpha}_{L}(n-\nu)g^{\alpha}_{L}(n) 
\ \ \ \ \ \ (\alpha = 0,x,y,z) \ ,
\label{defe}
\end{eqnarray}
\begin{eqnarray}
a^{(\pm)}_{\alpha}\!\! =\! g^{\alpha}_{L}(n\!-\!\nu)g^{0}_{L}\!(n)
\!\pm\! g^{0}_{L}(n\!-\!\nu)g^{\alpha}_{L}\!(n) \ \ (\!\alpha\! =\! x,\!y,\!z) ,
\label{defa}
\end{eqnarray}
\begin{eqnarray}
\hspace*{1mm}
b^{(\pm)}_{\alpha}\!\! =\! g^{\beta}_{L}(n\!-\!\nu)g^{\gamma}_{L}\!(n)
\!\pm\! g^{\gamma}_{L}(n\!-\!\nu)g^{\beta}_{L}\!(n) \ \  (\!\alpha\! =\! x,\!y,\!z) .
\label{defb}
\end{eqnarray}
Note that ($\alpha$, $\beta$, $\gamma$) in eq. (\ref{defb})
denotes a cyclic change of ($x, y, z$). 
The static Green functions $g^{\alpha}_{L}(n) \ (\alpha = 0,x,y,z)$ are
defined by $\tilde{g}_{L\sigma\sigma^{\prime}}(n)$ (see eq. (\ref{gst}))
as
\begin{eqnarray}
\tilde{g}_{L\sigma\sigma^{\prime}}(n) = 
g^{0}_{L}(n)\delta_{\sigma\sigma^{\prime}} + \sum_{\alpha}^{x,y,z} 
g^{\alpha}_{L}(n) (\sigma_{\alpha})_{\sigma\sigma^{\prime}} \ .
\label{defga}
\end{eqnarray}
\vspace{5mm}

\section{Calculation of the Gaussian average of dynamical potentials}
\vspace*{2mm}
We calculate here the Gaussian average of the $n$-th order products of
dynamical potentials.
\vspace{5mm}
\begin{eqnarray}
\overline{\Big[ \prod_{m=1}^{2l+1} \prod_{k=1}^{n(m)} 
(v_{\alpha_{k}(m)}(\nu,m)v_{\gamma_{k}(m)}(-\nu,m)) \Big]}
& = &  \nonumber \\ 
&  &  \hspace{-70mm}
\int
\Big[ \prod_{\alpha}^{xyz} \!\! \dfrac{\beta^{2l+1} {\rm det} 
B^{\alpha}}{(2\pi)^{2l+1}} \!\!
\prod_{m=1}^{2l+1} \!\!\!d^{2}\xi_{m\alpha}(\nu) \Big] 
\dfrac{\beta^{2l+1} {\rm det} A}{(2\pi)^{2l+1}}
\Big[ \!\!\prod_{m=1}^{2l+1} \!\!d^{2}\zeta_{m}(\nu) \Big] \ \ \ \ \ \ \ \ 
\nonumber \\
&  &  \hspace*{-65mm}
\times \ \Big[ \prod_{m=1}^{2l+1} \Big( \prod_{k=1}^{n(m)} 
v_{\alpha_{k}(m)}(\nu,m)v_{\gamma_{k}(m)}(-\nu,m) \Big) \Big]  
\nonumber \\
&  &  \hspace*{-65mm}
\times \, \exp \left[ - \frac{\beta}{2} 
\Big( 
\zeta^{\ast}(\nu) A \zeta(\nu) 
+ \sum_{\alpha} \xi^{\ast}_{\alpha}(\nu) B^{\alpha} \xi_{\alpha}(\nu)
\Big) \right] .  \hspace*{0mm}
\label{vvbar}
\end{eqnarray}
Here integers $\{ n(m) \}$ satisfy the constraint
$\sum_{m} n(m) = n$. 
$\zeta^{\ast}(\nu) A \zeta(\nu)$ stands for $\sum_{mm^{\prime}}
\zeta_{m}^{\ast}(\nu) A_{mm^{\prime}} \zeta_{m^{\prime}}(\nu)$. 
$v_{\alpha}(\nu,m)$ are given by eqs. (\ref{v0num}) and (\ref{vanum}).

The average is calculated from a generating function $I(s,t)$ as
follows. 
\vspace{3mm}
\begin{eqnarray}
\overline{\Big[ \prod_{m=1}^{2l+1} \prod_{k=1}^{n(m)} 
(v_{\alpha_{k}(m)}(\nu,m)v_{\gamma_{k}(m)}(-\nu,m)) \Big]} & = & 
\nonumber \\
&  &  \hspace{-75mm}
\Bigg[ \!\!\prod_{m=1}^{2l+1} \!\!\dfrac{\partial^{2n(m)}}
{\partial s_{\!m\!\alpha_{1}\!(m)} \partial t_{\!m\!\gamma_{1}\!(m)} \!\cdots\! 
\partial s_{\!m\!\alpha_{n(m)}\!(m)} \partial t_{\!m\!\gamma_{n(m)}\!(m)} }\!
\Bigg] \!I(\!s\!=\!0, \!t\!=\!0\!) . \hspace{8mm}
\label{vvbar2}
\end{eqnarray}
Here 
\begin{eqnarray}
I(s,t) \!\!\!\!\!\!& = & \!\!\!\!\!\!\!\!\int
\left[ \prod_{\alpha}^{xyz} \!\! \dfrac{\beta^{2l\!+\!1} \!{\rm det} 
B^{\alpha}}{(2\pi)^{2l\!+\!1}} \!\!\!\!
\prod_{m=1}^{2l+1} \!\!\!d^{2}\!\xi_{m\!\alpha}\!(\nu) \right]\!\!\! 
\dfrac{\beta^{2l\!+\!1} {\rm det} A}{(2\pi)^{2l\!+\!1}}
\left[ \prod_{m=1}^{2l+1} \!\!d^{2}\!\zeta_{m}\!(\nu) \right] 
\nonumber \\
&  &  \hspace*{-10mm}
\times \, \exp \Bigg[ - \frac{\beta}{2} 
\Big( 
\zeta^{\ast}(\nu) A \zeta(\nu) 
+ \sum_{\alpha} \xi^{\ast}_{\alpha}(\nu) B^{\alpha} \xi_{\alpha}(\nu)
\Big)   \nonumber \\
& &  \hspace*{0mm}
+ \sum_{m=1}^{2l+1} \sum_{\alpha=0}^{4}
\Big( s_{m\alpha}v_{\alpha}(\nu,m) + t_{m\alpha}v_{\alpha}(-\nu,m)
\Big) \Bigg] .
\label{ist}
\end{eqnarray}
The latter is obtained as follows.
\begin{eqnarray}
I(s,t) = \exp \Big[ \dfrac{1}{2\beta} \sum_{mm^{\prime}\alpha} s_{m\alpha} 
C^{\alpha}_{mm^{\prime}} t_{m^{\prime}\alpha}
\Big] \ .
\label{ist2}
\end{eqnarray}
Here the Coulomb interactions $C^{\alpha}_{mm^{\prime}}$ are 
defined by eq. (\ref{cdef}).

By differentiating $I(s,t)$ with respect to $s_{m\alpha}$
($t_{m\gamma}$), we have a new factor 
$(2\beta)^{-1} \sum_{n} C^{\alpha}_{mn} t_{n\alpha}$
($(2\beta)^{-1} \sum_{n} s_{n\gamma}C^{\gamma}_{nm}$).
When we take the $2n$-th derivative of $I(s,t)$ with respect to the
variable 
$(s_{m\alpha_{1}},t_{m\gamma_{1}}, \cdots , 
s_{m\alpha_{n}},t_{m\gamma_{n}})$ ,
we have a $2n$-th order polynomial times $I(s,t)$.
When we put $s_{m\alpha}=0$ and $t_{m\alpha}=0$ in the derivative, 
we have $I(s=0, t=0)=1$,
and only the 0-th order terms of the polynomial remain.  The latters were
created by taking a derivative of the factor $(2\beta)^{-1} \sum_{n}
s_{n\alpha} C^{\alpha}_{nm}$ or $(2\beta)^{-1} \sum_{n^{\prime}}
C^{\gamma}_{mn^{\prime}} t_{n^{\prime}\gamma}$ with respect to the
variable conjugate to $s_{m\gamma_{i}}$ or $t_{m^{\prime}\alpha_{i}}$.  
A created constant 
$(1/2\beta) C^{\alpha_{i}}_{mm^{\prime}}\delta_{\alpha_{i}\gamma_{j}}$ 
may be indicated by a contraction 
$\overline{s_{m\alpha_{i}}t_{m^{\prime}\gamma_{j}}}$.
Then the 0-th order terms, and therefore the Gaussian average
(\ref{vvbar}) should be given by the sum over all possible
products of contractions.
\begin{eqnarray}
\overline{\Big[ \prod_{m=1}^{2l+1} \prod_{k=1}^{n(m)} 
(v_{\alpha_{k}(m)}(\nu,m)v_{\gamma_{k}(m)}(-\nu,m)) \Big]} &=&
\nonumber \\
& &  \hspace{-65mm}
\dfrac{1}{(2\beta)^{n}} \sum_{\rm P}
\Bigg[ \prod_{m=1}^{2l+1} \prod_{k=1}^{n(m)} 
C^{\alpha_{k}(m)}_{mm_{\rm p}} 
\delta_{\alpha_{k}(m)\gamma_{k_{\rm p}}(m_{\rm p})}
\Bigg] .\hspace{5mm}
\label{vvbar3}
\end{eqnarray}
Here the permutation P is taken with respect to the $n$ elements 
$\{ (k,m) \, | \, k=1, \cdots, n(m); \, m=1, \cdots, 2l+1 \}$; 
${\rm P}\{ (k, m) \} = \{ (k_{\rm p}, m_{\rm p})\}$.
Application of the formula (\ref{vvbar3}) to eq. (\ref{dnubar}) yields
eq. (\ref{dnubarn}) in \S 4:
\begin{eqnarray}
\overline{D}^{(n)}_{\nu}\!\!\!\! &=&\!\!\!\! \dfrac{1}{(2\beta)^{n}} 
\sum_{\sum_{km} l(k,m)=n} \sum_{\{ \alpha_{j}(k,m)\} }
\sum_{\rm P}     \nonumber \\
& &  \hspace{-5mm}
\prod_{m=1}^{2l+1} \prod_{k=0}^{\nu-1}
\Bigg[ \Big( \prod_{j=1}^{l(k,m)} C^{\alpha_{j}}_{mm_{\rm p}} \Big)
\hat{D}^{(l(k,m))}_{\{ \alpha\alpha_{{\rm p}^{-1}} \} }(\nu,k,m) \Bigg]
.  \hspace{7mm}
\label{dnubarn2}
\end{eqnarray}


\begin{thebibliography}{99} 
%
%
%
\bibitem{fulde95} See for example, P. Fulde: {\it Electron Correlations
	in Molecules and Solids} (Springer Verlag. Pub., Berlin, 1995).
\bibitem{imada98} M. Imada, A. Fujimori, and Y. Tokura:
	Rev. Mod. Phys. {\bf 70} (1998) 1039.
\bibitem{kake04} Y. Kakehashi: Adv. in Phys. {\bf 53} (2004);
	Phil. Mag. {\bf 86} (2006) 2603.
\bibitem{fulde06} P. Fulde, P. Thalmeier, G. Zwicknagl, Solid State
	Phys. {\rm 60} (2006) 1.
\bibitem{eastman80} D.E. Eastman, F.J. Himpsel, and J.A. Knapp,
	Phys. Rev. Lett. {\bf 14} (1980) 95.
\bibitem{schafer05} J. Sch\"afer, M. Hoinkis, Eli Rotenberg, P. Blaha,
	and R. Claessen, Phys. Rev. B {\bf 72} (2005) 155115.
\bibitem{bozorth68} R.M. Bozorth: {\it Ferromagnetism} (Van Nostrand,
	Princeton, 1968).
\bibitem{himpsel79} F.J. Himpsel, J.A. Knapp, and D.E. Eastman: Phys. B
	{\bf 19} (1979) 2919.
\bibitem{moruzzi95} See for example, V.L. Moruzzi and C.B. Sommers: {\it
	Calculated Electronic Properties of Ordered Alloys: A Handbook}
	(World Scientific Pub., Singapore, 1995).
\bibitem{moruzzi78} See for example, V.L. Moruzzi, J.F. Janak, and
	A.R. Williams: {\it Calculated Electronic Properties of Metals}
	(Pergamon, New York, 1978).
\bibitem{cyrot72} M. Cyrot: J. Phys. (Paris) {\bf 33} (1972) 25.
\bibitem{hub79} J. Hubbard: Phys. Rev. B {\bf 19} (1979) 26267; B {\bf 20}
	(1979) 4584; B {\bf 23} (1981) 5974.
\bibitem{hase79} H. Hasegawa: J. Phys. Soc. Jpn {\bf 46} (1979)
	1504; {\bf 49} (1980) 178.
\bibitem{gutz63} M.C. Gutzwiller: Phys. Rev. Lett. {\bf 10} (1963) 159;
	Phys. Rev. {\bf 134} (1964) A293; Phys. Rev. {\bf 137} (1965)
	A1726.
\bibitem{hub63} J. Hubbard: Proc. Roy. Soc. (London) A {\bf 276} (1963) 
        238; A {\bf 281} (1964) 401.
\bibitem{kana63} J. Kanamori: Prog. Theor. Phys. {\bf 30} (1963) 275.
\bibitem{kake92} Y. Kakehashi: Phys. Rev. B. {\bf 45} (1992) 7196;
	J. Magn. Magn. Mater. {\bf 104-107} (1992) 677.
\bibitem{kake02} Y. Kakehashi: Phys. Rev. B {\bf 65} (2002) 184420.
\bibitem{amit71} D.J. Amit and C.M. Bender: Phys. Rev. B {\bf 4} (1971)
	3115; D.J. Amit and H.J. Keiter: Low Temp. Phys. {\bf 11} (1973)
	603.
\bibitem{dai91} Dai Xianxi: J. Phys. Condens. Matter. {\bf 3} (1991)
	4389.
\bibitem{ander75} O.K. Andersen: Phys. Rev. B {\bf 12} (1975).
\bibitem{ander94} O.K. Andersen, O. Jepsen, and G. Krier: in {\it
	Methods of Electronic Structure Calculations} ed. by V. Kumar,
	O.K. Andersen, and A. Mookerjee (World Scientific Pub.,
	Singapore, 1994) p. 63.
\bibitem{par89} See for example, R. G. Parr and W. Yang:
	{\it Density Functional Theory of Atoms and Molecules} (Oxford
	University Press., Oxford, 1989).
\bibitem{barth72} U. von Barth and L. Hedin: J. Phys. C {\bf 5} (1972)
	1629.
\bibitem{georges96} A. Georges, G. Kotliar, W. Krauth, M.J. Rosenberg:
	Rev. Mod. Phys. {\bf 68} (1996) 13.
\bibitem{kake02-2} Y. Kakehashi: Phys. Rev. B {\bf 66} (2002) 104428.
\bibitem{anisimov97} V.I. Anisimov, F. Aryasetiwan, and
	A.I. Lichtenstein: J. Phys. Condens. Matter {\bf 9} (1997) 767.
\bibitem{hirsch89} J.E. Hirsch and R.M. Fye: Phys. Rev. Lett. {\bf 56}
	(1989) 2521.
\bibitem{lich01} A.I. Lichtenstein and M.I. Katsnelson, and G. Kotliar:
	Phys. Rev. Lett. {\bf 87} (2001) 067205.
\bibitem{vidberg77} H.J. Vidberg and J.W. Serene: J. Low
	Temp. Phys. {\bf 29} (1977) 179.
\bibitem{morandi74} See for example, G. Morandi, E. Galleani, D'Agliano,
	F. Napoli, and C.F. Ratto: Adv. Phys. {\bf 24} (1974) 867.
\bibitem{hub59} J. Hubbard: Phys. Rev. Lett. {\bf 3} (1959) 77;
	R.L. Stratonovich: Dokl. Akad. Nauk. SSSR {\bf 115} (1958) 1097;
	Sov. Phys. Dokl. {\bf 2} (1958) 416.
\bibitem{ehren76} See for example, H. Ehrenreich and L.M. Schwartz:
	Solid State Phys. {\bf 31} (1976) 1.
\bibitem{anisimov97-2} V.I. Anisimov, A.I. Poteryaev, M.A. Korotin,
	A.O. Anokhin, and G. Kotliar: J. Phys. Condens. Matter {\bf 9}
	(1997) 7359.
\bibitem{anisimov93} V.I. Anisimov, I.V. Solovyev, and M.A. Korotin,
	M.T. Czy\.{z}yk, and G.A. Sawtzky: Phys. Rev. B {\bf 48} (1993)
	16929.
\bibitem{kirk70} S. Kirkpatrick, B. Velick\'{y}, and H. Ehrenreich:
	Phys. Rev. B {\bf 1} (1970) 3250. 
\bibitem{unger94} P. Unger, J. Igarashi, and P. Fulde: Phys. Rev. B {\bf
	50} (1994) 10485.
\bibitem{drchal99} V. Drchal, V. Jani\u{s}, and J. Kudrnovsk\'{y}:
	Phys. Rev. B {\bf 60} (1999) 15664.
\bibitem{fallot44} M. Fallot: J. de Phys. Rad. V (1944) 153.
\bibitem{arrot67} A. Arrott and J.E. Noakes: Phys. Rev. Lett. {\bf 19}
	(1967) 786.
\bibitem{danan68} H. Danan, A. Herr, and A.J.P. Meyer:
	J. Appl. Phys. {\bf 39} (1968) 669.
\bibitem{potter34} H.H. Potter: Proc. Roy. Soc. London A {\bf 146} (1934)
	S362. 
\bibitem{sucksmith38} W. Sucksmith and R.R. Pearce:
	Proc. Roy. Soc. (London) A {167} (1938) 189.

\end{thebibliography}
\end{document}